\documentclass[%
 aip,
 amsmath,amssymb,
 reprint,%
]{revtex4-1}

\usepackage{graphicx}
\usepackage{dcolumn}
\usepackage{bm}

\usepackage[utf8]{inputenc}
\usepackage[T1]{fontenc}
\usepackage{mathptmx}
\usepackage{etoolbox}
\usepackage{verbatim}
\usepackage{hyperref}
\usepackage{bbold}
\usepackage{xcolor}
\usepackage{mathrsfs}

\makeatletter
\def\@email#1#2{%
 \endgroup
 \patchcmd{\titleblock@produce}
  {\frontmatter@RRAPformat}
  {\frontmatter@RRAPformat{\produce@RRAP{*#1\href{mailto:#2}{#2}}}\frontmatter@RRAPformat}
  {}{}
}%
\makeatother
\begin{document}

\preprint{AIP/123-QED}

\title[]{Achiral dipoles on a ferromagnet can affect its magnetization direction}

\author{Ragheed Alhyder}
 \affiliation{Institute of Science and Technology Austria (ISTA), \\
 Am Campus 1, 3400 Klosterneuburg, Austria}

\author{Alberto Cappellaro}
 \email{alberto.cappellaro@ist.ac.at}
 \affiliation{Institute of Science and Technology Austria (ISTA), \\
 Am Campus 1, 3400 Klosterneuburg, Austria}
 
\author{Mikhail Lemeshko}%
 \affiliation{Institute of Science and Technology Austria (ISTA), \\
 Am Campus 1, 3400 Klosterneuburg, Austria}

\author{Artem G. Volosniev}
 \affiliation{Institute of Science and Technology Austria (ISTA), \\
 Am Campus 1, 3400 Klosterneuburg, Austria}

\date{\today}

\begin{abstract}
We demonstrate the possibility of a coupling between the magnetization direction of a ferromagnet and the tilting angle of adsorbed achiral molecules. To illustrate the mechanism of the coupling, we analyze a minimal Stoner model that includes Rashba spin-orbit coupling due to the electric field on the surface of the ferromagnet.
The proposed mechanism allows us to study magnetic anisotropy of the system with an extended Stoner-Wohlfarth model, and argue that adsorbed achiral molecules can change magnetocrystalline
anisotropy of the substrate.
Our research's aim is to motivate further experimental studies of the current-free chirality induced spin selectivity effect involving {\it both} enantiomers. 
\end{abstract}

\maketitle

\section{Introduction}

Chirality induced spin selectivity (CISS) has become an umbrella name for a number of seemingly related experimental reports on the coupling between geometrical chirality 
and the spin degree of freedom~\cite{Naaman2019,Waldeck2021,Naaman2022,Xu2023}. 
Initially, the effect was discovered by analyzing photo-excited electrons after they pass through adsorbed biological molecules~\cite{Ray1999,Ghler2011} and in currents of electrons through organic chiral junctions~\cite{Xie2011}. Later, the interplay between chirality and the electron's spin was observed in chemical reactions~\cite{Zhang2018,Metzger2020}, in enantioselective adsorption on magnetic substrates~\cite{BanerjeeGhosh2018,safari2022enantiospecific,Santra2023}, and even in fully inorganic systems~\cite{Ghosh2019,Inui2020}. By now, there is a great collection of experimental set-ups and observables that demonstrate CISS. In spite of this, detailed theoretical understanding of the CISS effect, much needed for a further development of the field, is still lacking~\cite{Evers2022,Fransson2022}.

One problem in building theoretical models of CISS is that the experimental platforms differ drastically from one another, and are traditionally studied using independent methods and approaches.
This does not rule out the possibility of a unifying  mechanism behind the CISS effect; it suggests, however, to study the existing classes of experiments separately. From the theoretical standpoint, it is natural to start by examining systems where the interplay between spin and chirality is observed without applied currents~\cite{Sukenik2020,Miwa_2020,Meirzada2021,Alpern2021},
note also relevant theoretical models in Refs.~\onlinecite{Volosniev2021,fransson-nano-2021,fransson-jpcl-2022}.
First, these systems are usually easier to analyze, as they can be studied using methods developed for steady states. Second, looking into these systems might help to answer fundamental questions about the  CISS effect, in particular, about the origin of time-reversal-breaking correlations necessary for spin-selective phenomena. 
Indeed, systems in equilibrium have vanishing currents, eliminating  the key ingredient of many CISS-related theoretical models~\cite{Gutierrez2012,Guo2012,Medina2012,Gersten2013,Guo2014,Matityahu2016,Diaz_2018,Fransson2019,Michaeli2019,Dalum2019,Fransson2020,Ghazaryan2020,Ghazaryan2020a,Alwan2021,Liu2021,Wolf2022,Vittmann2023} from the picture. In the absence of currents, time reversal breaking may occur, e.g., as a result of substrate magnetization~\cite{Sanvito2010} or of non-unitary effects (such as dephasing and  dissipation) that are used in CISS models~\cite{Guo2012,Guo2014,Matityahu2016,Geyer2019,Volosniev2021}.

\begin{figure}
\includegraphics[scale=0.8]{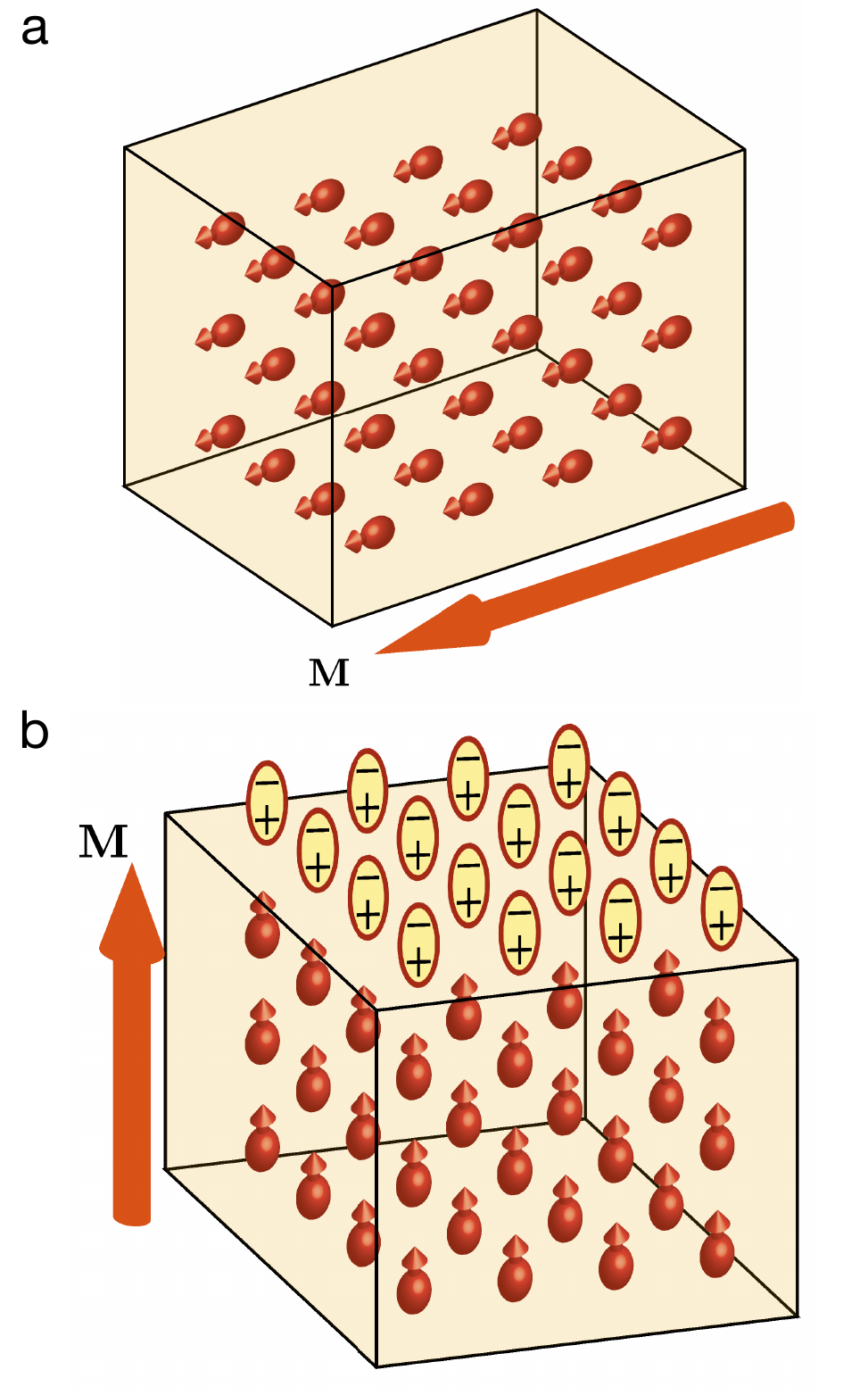}
\caption{Molecules adsorbed on a magnet modify surface electric fields,
which, in turn, can change magnetocrystalline anisotropy of the substrate. 
Panel a (b) illustrates the preferred magnetization direction without (with) a molecular self-assembled layer. The molecules are shown as dipoles, see panel~b. 
Spin-polarized electrons inside the substrate are presented as balls with arrows. Their direction determines the direction of magnetization, $\mathbf{M}$.
}
\label{fig:key_message}
\end{figure}

Arguably, the most puzzling observation of  current-free CISS experiments
is the coupling between the direction of magnetization in the substrate and the tilt of the organic molecules~\cite{Sukenik2020,Meirzada2021}. 
Although this observation is typically explained within the CISS framework, in this contribution we want to highlight the fact that the magnetization of the substrate, $\mathbf{M}$, can in principle couple directly to the polarization of the molecular layer, $\mathbf{P}$, without the need to involve chirality. For example, a phenomenological term in the form $\beta (\mathbf{P}\cdot \mathbf{M})^2$ can enter the energy functional since it is allowed by the time-reversal symmetry of the problem.
Here, the parameter $\beta$ determines the strength of the effect.
Some experimental results~\cite{Sukenik2020, Meirzada2021} can be explained by the term $\beta (\mathbf{P}\cdot \mathbf{M})^2$ if $\beta$ is negative so that the magnetization prefers to be parallel to the direction of $\mathbf{P}$, see Fig.~\ref{fig:key_message}.
For example, this explains the larger tilting angle for samples with the in-plane easy axis than for the out-of-plane easy axis~\cite{Sukenik2020}, and also the correlation between the direction of the magnetization and the tilt angle of the molecular layer~\cite{Meirzada2021}.

 As we shall illustrate in this paper, $\beta (\mathbf{P}\cdot \mathbf{M})^2$ is not the only term allowed by symmetries, and a case-by-case study is probably needed to establish accurate phenomenological models for every system of interest. For example, on the surface of a metal only the normal component of the electric field should  play a role, so that the terms of the form $\beta(\mathbf{P}\cdot\mathbf{n})^2(\mathbf{M}\cdot\mathbf{n})^2$ 
 become relevant, where $\mathbf{n}$ is the vector normal to the surface.
 In any case, the very possibility of a chirality-independent coupling between the magnetization direction and a tilting angle must motivate  current-free experimental studies of the CISS effect involving {\it both} enantiomers.
This might be especially important in light of the observation that, for certain substrates, the effect of chirality on the strength of the coercive field might be a next-to-leading-order effect~\cite{Miwa_2020}.

\section{Magneto-electric coupling}

The key message of our study is that the electric field at the interface between a ferromagnet and a molecular layer can couple to the vector $\mathbf{M}$, see Fig.~\ref{fig:key_message}. This hypothesis falls into the realm of  magneto-electric coupling (MEC) effects in a broad sense~\cite{Eerenstein2006}. Many properties of MEC are well-established, therefore, we find it appropriate to review them here only briefly.

It is expected that the interface MEC is very sensitive to the electronic and atomic structure of constituents~\cite{Niranjan2010,KUMARI2021}, in particular, because reactive metal surfaces of ferromagnetic materials may lead to formation of chemical bonds between the molecules and the substrate~\cite{Cinchetti2017}.
Therefore, a microscopic first-principle calculation is required for every experimental set-up to accurately determine the resulting MEC. Here,     
we do not attempt such a calculation, as our goal is to provide a basic intuition for phenomenological terms that can couple the directions of $\mathbf{P}$ and $\mathbf{M}$.  In particular, we illustrate the effect using a toy model that explains a physical mechanism of coupling, and symmetries of the problem. 

The mechanism connecting electric and magnetic effects in our case is Rashba spin-orbit coupling (SOC), which is known to be important at surfaces and interfaces~\cite{
Soumyanarayanan2016}.  
Before introducing the toy model in the following section, we note that 
there are a number of other effects that can lead to the interplay between magnetization of the surface and adsorbed molecules. For example, the 
adsorption of molecules can increase the strength of the magnetic exchange interaction on the surface~\cite{callsen2013} and even invert the magnetic anisotropy of thin films under certain conditions~\cite{Bairagi2015}. 
These `magnetic hardening' effects rely on chemical bonding between molecular adsorbates and the substrate, and cannot directly explain the experiments where the ferromagnet is coated with gold\cite{Meirzada2021}.  
Screening charge on the surface of a ferromagnetic metal can also modify magnetization of the surface~\cite{Zhang1999, Niranjan2010}. However, in this effect the direction of $\mathbf{M}$ is not important, and therefore it cannot directly explain the experimental data taken for different directions of magnetization\cite{Sukenik2020}. Finally, we note that 
density-functional theory calculations on
isolated ferromagnetic films indicate that external electric fields can be used to change the magnetization between in-plane and out-of-plane orientations~\cite{Duan2008}. This observation is in line with our phenomenological results, see below.  


\section{Electrostatics of the molecular layer}

{\bf Formulation.} The simplest approach to static properties of electrons at interfaces is based upon electrostatics~\cite{Natan2007,Monti2012}. In this approach, the main effect of the organic layer is in the change of the electric field across the interface. For simplicity, we assume that the electric field on the substrate is determined completely by the molecules. Furthermore, as we are interested in surface effects, we restrict the motion of electrons to the $xy$ plane. With these assumptions, we write the following Stoner model (cf.~Ref.~\onlinecite{Gold1974}), 
\begin{equation}
\hat{H}_{\text{tot}} = \epsilon_{\mathbf{k}}\mathbb{I}_{2} + \mathcal{P} \mathbf{I}\cdot \pmb{\sigma} +  \hat{H}_{\text{SOC}}\;,
\label{total hamiltonian single electron}
\end{equation}
where $\mathbb{I}_{2}$ is the $2\times 2$ unity matrix, and $\epsilon_{\mathbf{k}}$ is 
a dispersion relation without SOC and Stoner terms.  For simplicity, we assume that $\epsilon_{\mathbf{k}}$ is independent of spin.
$\mathcal{P} \mathbf{I}\cdot \pmb{\sigma}$ is the Stoner term, which we introduce to account for a ferromagnetic nature of the substrate; here, $\pmb{\sigma}$ is the spin operator, $I=|\mathbf{I}|$ is the Stoner parameter and $\mathcal{P}$ is the polarization of the substrate, see below. 
If $\hat{H}_{\text{SOC}}$ is vanishing, then the direction of $\mathbf{I}$ defines the direction of magnetization, and a natural quantization axis for the spin with projections: $\uparrow_I$ and $\downarrow_I$. The corresponding electronic densities are $n_{\uparrow_I}=N_{\uparrow_I}/S$ and $n_{\downarrow_I}=N_{\downarrow_I}/S$, where $N_{\uparrow_I/\downarrow_I}$ is the number of particles and $S$ is the surface area.  The corresponding polarization of the substrate reads $\mathcal{P}=(n_{\uparrow_I}-n_{\downarrow_I})/(n_{\uparrow_I}+n_{\downarrow_I})$. Without loss of generality, we assume that $\mathbf{I} = I_x \mathbf{e}_x 
+ I_z \mathbf{e}_z$ (thus $I_y=0$).

The last term in Eq.~\eqref{total hamiltonian single electron} describes a Rashba-like spin-orbit coupling
\cite{Rashba1960,rashba-1984,manchon-2015}
\begin{equation}
\hat{H}_{\text{SOC}} = \alpha_R \, \pmb{\sigma} \cdot \big(\mathbf{E} \times \hat{\mathbf{p}} \big)\;,
\label{rashba soc single electron}
\end{equation}
which couples the electric field to electron's spin, providing the basis for MEC.
$\alpha_R$ is a constant that determines the strength of SOC;  $\mathbf{E}$ is a homogeneous (at least in first approximation) electric 
field, originating from the charge re-organization concurrent to the chemical absorption process for
the molecular monolayer. 

Since the substrate is metallic, the electric field has to be perpendicular to its surface, $\mathbf{E} = E_z \mathbf{e}_z$. We assume that the component $E_z$ is proportional to $P_z=P\cos(\theta_P)$, and its amplitude is therefore determined by the tilt angle of the molecular layer, $\theta_P$, see Fig.~\ref{fig:app} in the Appendix. The assumption $E_z\sim P_z$ is not essential, and used only to simplify the discussion. It is only important that by tilting the angle of the molecules, the value of $E_z$ changes. 
In general, the direction of $\mathbf{I}$ is not perpendicular to the surface, and our goal here is to investigate the energy as a function of $E_z$ and the direction of~$\mathbf{I}$. 

For the sake of completeness, it is worth remarking that one must exercise care when 
working with Eq. \eqref{rashba soc single electron}, since hermiticity of the resulting Hamiltonian is not guaranteed for at least a couple of relevant situations.
The first scenario, now well understood in solid-state physics, involves the dimensional
reduction, occuring when, for instance, a strong potential confines electrons
on an effective one-dimensional ring (see Refs. \onlinecite{meijer-2002} and \onlinecite{berche-2010}
for an extensive discussion about this issue).
Second, note that spin-orbit coupling is an inherently relativistic effect as its
derivation might be based upon the Dirac theory\cite{bransden-quantum-book}. 
The lowest-order relativistic corrections to the Hamiltonian are given
\cite{bercioux-2015,mondal-2015,mondal-2016,Mondal-2017}
by Eq.~\eqref{rashba soc single electron} and one more term, interpreted as the
coupling between the angular momentum of a radiation field and the magnetic 
moment of the electron. 
For our purpose, Eq.~\eqref{rashba soc single electron} is accurate
because the electric field is assumed to be time- and position-independent and the relativistic corrections to the eigenstates are neglected.

{\bf Eigenstates of the Hamiltonian.} To elucidate the interplay between $\mathbf{I}$ and the Rashba SOC, we calculate the spectrum
of $\hat{H}_{\text{tot}}$ from Eq.~\eqref{total hamiltonian single electron}. To this end, we work in the momentum representation, where the total Hamiltonian is cast in the form 
\begin{equation}
\begin{aligned}
\hat{H}_{\text{tot}} & = \epsilon_{\mathbf{k}} \mathbb{I}_{2}\;-  \mathcal{P} I_z \sigma_z -\alpha_R E_z  \sigma_y k_x \\
&\qquad \qquad + \sigma_x \bigg( \alpha_R E_z k_y - \mathcal{P}I_x\bigg) \;.
\end{aligned}
\label{total hamiltonian momentum representation}
\end{equation}
 Note that the Pauli matrices anticommute with each other, and therefore, we can easily calculate the spectrum by considering $(\hat{H}_{\text{tot}}-\epsilon_{\mathbf{k}}\;\mathbb{I}_{2})^2$. We derive
\begin{equation}
E_{\mathbf{k}s} = \epsilon_{\mathbf{k}} - 
s \sqrt{\alpha_R^2 E^2_z k_x^2 + \mathcal{P} I_z^2  + \bigg(\alpha_R E_z k_y - \mathcal{P} I_x\bigg)^2},
\label{spectrum bands constant fields}
\end{equation}
where $s= \pm 1$ labels the spin manifold.

\begin{figure}
\includegraphics[width=0.9\columnwidth]{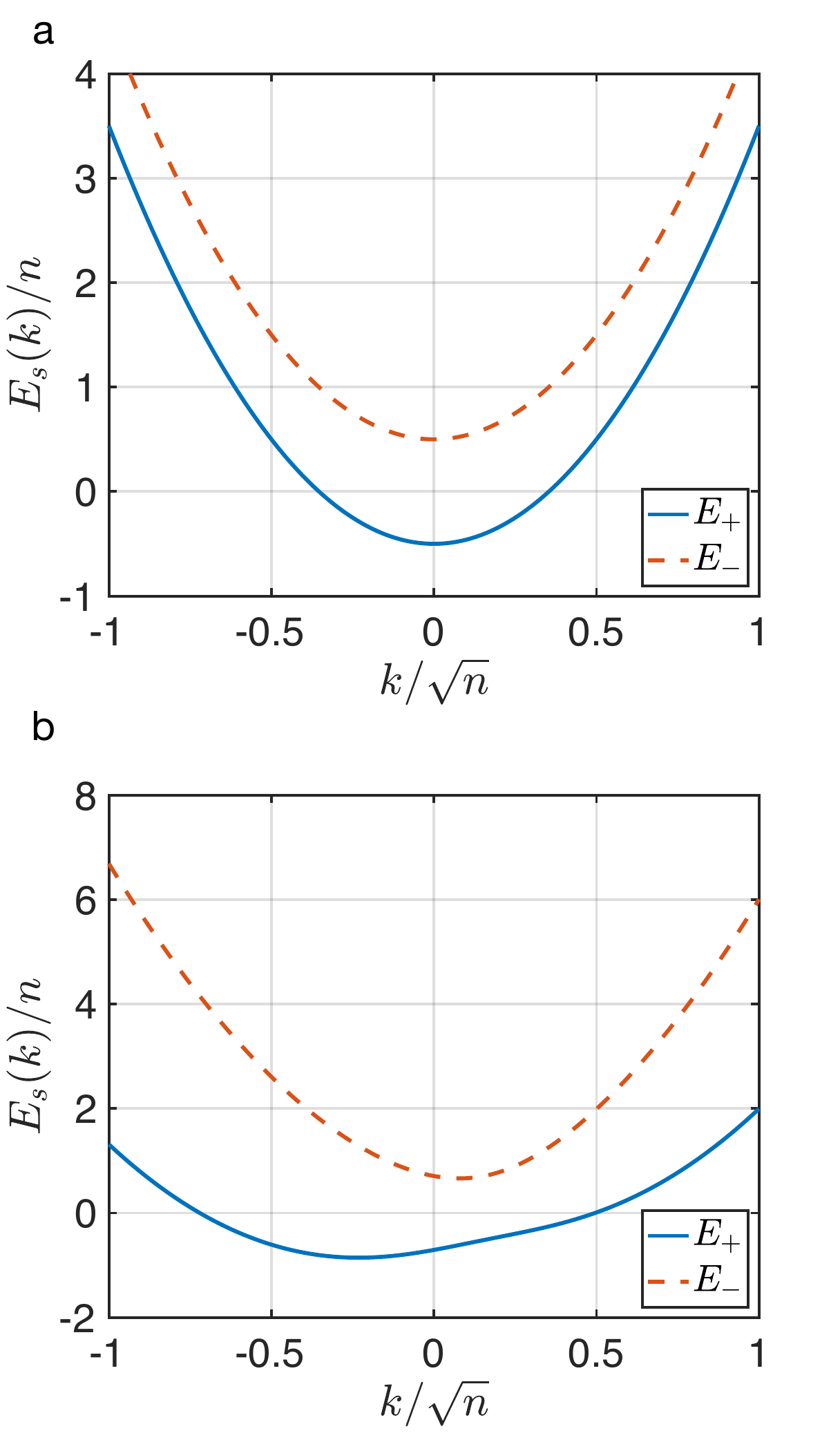}
\caption{Energy spectrum of the Hamiltonian $\hat{H}_{\text{tot}}$ as given by  Eq.~\eqref{spectrum bands constant fields}. The energy is plotted for $k_x=k_y \equiv k$.  The parameter $1/\sqrt{n}$ is used as a unit of length ($n$ is the density of electrons in the substrate). 
a: the spectrum without the electric field, the parameters are $\mathcal{P} I_z = 0.25$, $\mathcal{P} I_x = 0$ and $E_z = 0$. b: the spectrum for a non-vanishing electric field, the parameters are $\mathcal{P} I_z = 0.25$, $\mathcal{P} I_x = 0$ and $\alpha_R E_z = 0.8$ (with the
convention $\hbar \equiv m\equiv n \equiv 1$).
} 
\label{fig:1}
\end{figure}
In Fig.~\ref{fig:1} we plot the energy spectrum bands as provided by the equation above for $\epsilon_{\mathbf{k}}=k^2/2$ (here, $\hbar\equiv m\equiv1$).
As a matter of convention, the lower bands have been labelled as
$E_{+}(\mathbf{k})$. By construction, the Stoner term  introduces a gap between the $s = \pm 1$ bands,
while the Rashba SOC (cf. Eq.~\eqref{rashba soc single electron}) shifts the band minima.

{\bf Total energy of the system.} The total energy of the system
is obtained by summing over the lowest $N_{\uparrow_I}+N_{\downarrow_I}$ states $|\mathbf{k}s\rangle$, namely 
\begin{equation}
E^{\text{(tot)}} = \sum E_{\mathbf{k}s}\;,
\label{total energy definition}
\end{equation}
 where the sum is most easily calculated 
 in the continuum limit: $\sum_{\mathbf{k}} \rightarrow \frac{S}{(2\pi)^2}\int 
d^2\mathbf{k}$. 

\begin{figure}
 \includegraphics[scale=0.55]{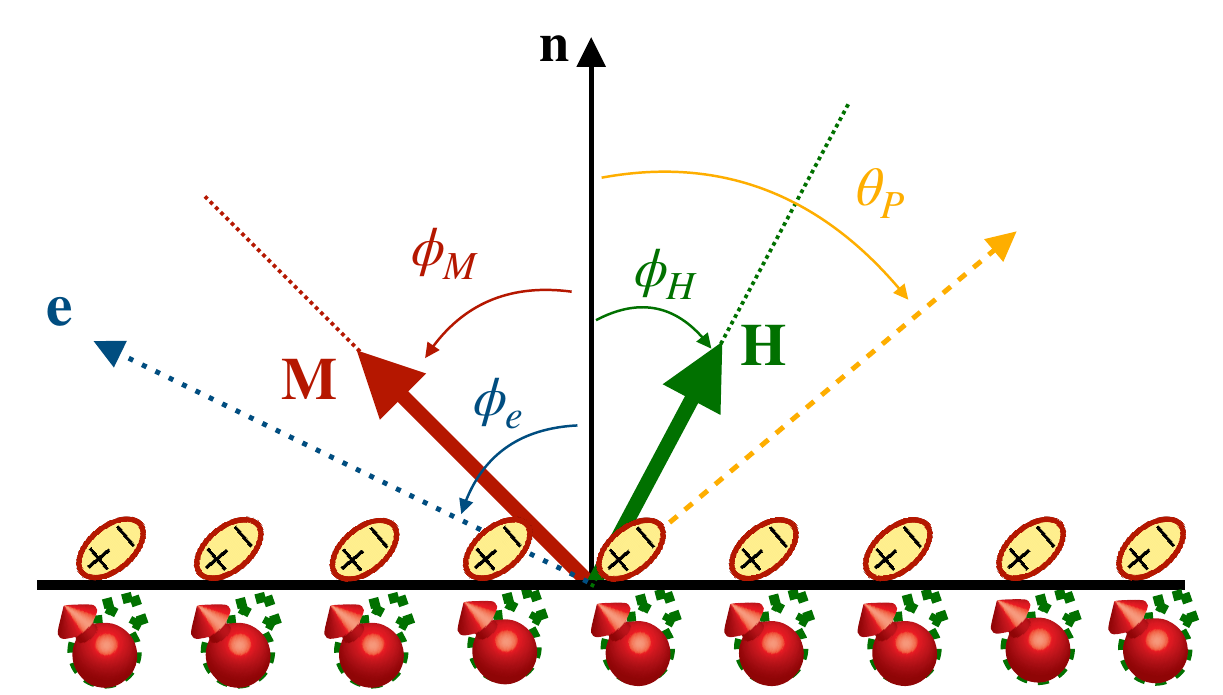}
 \caption{The geometry of the extended Stoner-Wohlfarth model. The angles are measured with respect to the normal to the substrate, $\mathbf{n}$.  The figure shows the directions of magnetization, $\mathbf{M}$; external magnetic field, $\mathbf{H}$; easy axis, $\mathbf{e}$; also shown is the tilting angle of the dipoles, $\theta_P$. The direction of an arrow below each angle determines the sign of the angle, e.g., in our convention $\phi_M$ is negative, whereas $\phi_H$ is positive.
 }
 \label{fig:geometry}
 \end{figure}

To illustrate this energy, we assume that the energy scale associated with SOC is significantly
smaller than all other energy scales, i.e., $E_z\to0$. With this assumption, we write the energy up to $E_z^2$-terms as 
\begin{align}
E_{\mathbf{k}s} \simeq  \epsilon_{\mathbf{k}} - 
s  \bigg(\mathcal{P}I-\alpha_R E_z k_y \sin(\theta_I)+\nonumber \\
\frac{\alpha_R^2E_z^2k^2}{2\mathcal{P}I}-\frac{\alpha_R^2 E_z^2k_y^2\sin^2(\theta_I)}{2\mathcal{P}I}\bigg),
\label{eq:Eks_expansion}
\end{align}
where $\sin(\theta_I)=I_x/I$. 
To demonstrate that the change in the energy due to the presence of the organic layer 
\begin{equation}
\Delta E=E^{\text{(tot)}}(E_z)-E^{\text{(tot)}}(E_z=0),
\end{equation}
can depend on the angle $\theta_I$, we assume that $\mathcal{P}=1$ and $I\to \infty$, i.e., we work with a strong ferromagnet. In this case, only the first line in Eq.~(\ref{eq:Eks_expansion}) is relevant because there is a high energy price for flipping a spin of an electron.  We derive
\begin{equation}
\frac{1}{S}\frac{\Delta E}{ \mathcal{E}_F}\simeq  - \sin^2(\theta_I)
\frac{m^2 \alpha_R^2E_z^2}{4\pi},
\label{eq:energy_shift}
\end{equation}
where $\mathcal{E}_F$ is the Fermi energy of the `spin-up' particles. 
The dependence of $\Delta E / \mathcal{E}_F$ on $\theta_{I}$
 shows that the system prefers to have in-plane magnetization. Note that Eq.~(\ref{eq:energy_shift}) contradicts the sketch in Fig.~\ref{fig:geometry}. However, this is not a point of concern for two reasons. First, our goal here is to illustrate general principles of coupling, and the overall sign of the derived term may change if other toy models are considered. Second, the surface electric field $E_z$ exists in real materials even without the molecules. Therefore, the following representation of $E_z$ might be more accurate: $E_z\simeq E_z^0+\cos(\theta_P) E_z^1$. If $E_z^0$ and $E_z^1$ have different signs and $|E_z^0|\gg |E_z^1|$, the situation in Fig.~\ref{fig:geometry} is restored. The most important conclusion of this section is that  
 the strength of the effect presented in Eq.~(\ref{eq:energy_shift}) is  determined by the tilt angle of the molecules via $E_z$.

\section{Extended Stoner-Wohlfarth model}

The discussion in the previous section provides a basis for studying magneto-electric coupling between organic molecules and the ferromagnetic substrate. With it, we can extend the simple Stoner-Wohlfarth model~\cite{Stoner1948, Tannous_2008} and understand the effect of the molecules on magnetic hysteresis and the interplay between the molecular tilting angle and the direction of $\mathbf{M}$.  Assuming that the ferromagnet has an uniaxial magnetic anisotropy, we write the energy, $\mathcal{E}$, as
\begin{align}
\frac{\mathcal{E}}{K_u \mathcal{V}}=&\sin^2(\phi_M-\phi_e)-\frac{\mu_0 M_s H}{K_u} \cos(\phi_M-\phi_H) \nonumber \\ &-\frac{\alpha \cos^2(\theta_P) }{K_u d} \sin^2(\phi_M)
\label{extended stoner model}
\end{align}
where the first line is the standard Stoner-Wohlfarth model, and the second line is our extension. Here, $\mathcal{V}$ is the three-dimensional volume of the ferromagnet; $K_u$ is the anisotropy parameter; $M_s$ is the saturation magnetization; $H$ is the magnitude of an external magnetic field; $\mu_0$ is the vacuum permeability. The last term in Eq.~(\ref{extended stoner model}) is derived in Eq.~\ref{eq:energy_shift}. It introduces magnetic anisotropy due to the presence of organic molecules. Its strength is determined by
\begin{equation}
\alpha=\frac{\mathcal{E}_F\alpha_R^2E_z^2m^2}{4\pi \cos^2(\theta_P)},
\end{equation} 
where the parameter $d$ is the thickness of the ferromagnet. In what follows, we shall treat the amplitude and the sign of $\alpha$ as free parameters, and estimate `realistic' values of $|\alpha|$ only at the end of this section.
Finally, we note that we associate the angle of the magnetization with that of~$\mathbf{I}$, i.e.,~$\phi_M=\theta_I$.   

 The geometry of the system is presented in Fig.~\ref{fig:geometry}. 
We measure all angles with respect to the normal to the surface. $\phi_M$, $\phi_H$, and $\phi_e$ define the angles of the magnetization, external magnetic field and the easy axis, respectively.
We assume that $\mathbf{H}, \mathbf{e}, \mathbf{n}$ and $\mathbf{M}$ all lie in one plane. This assumption does not change the qualitative nature of our discussion.

\begin{figure}
\includegraphics[scale=0.45]{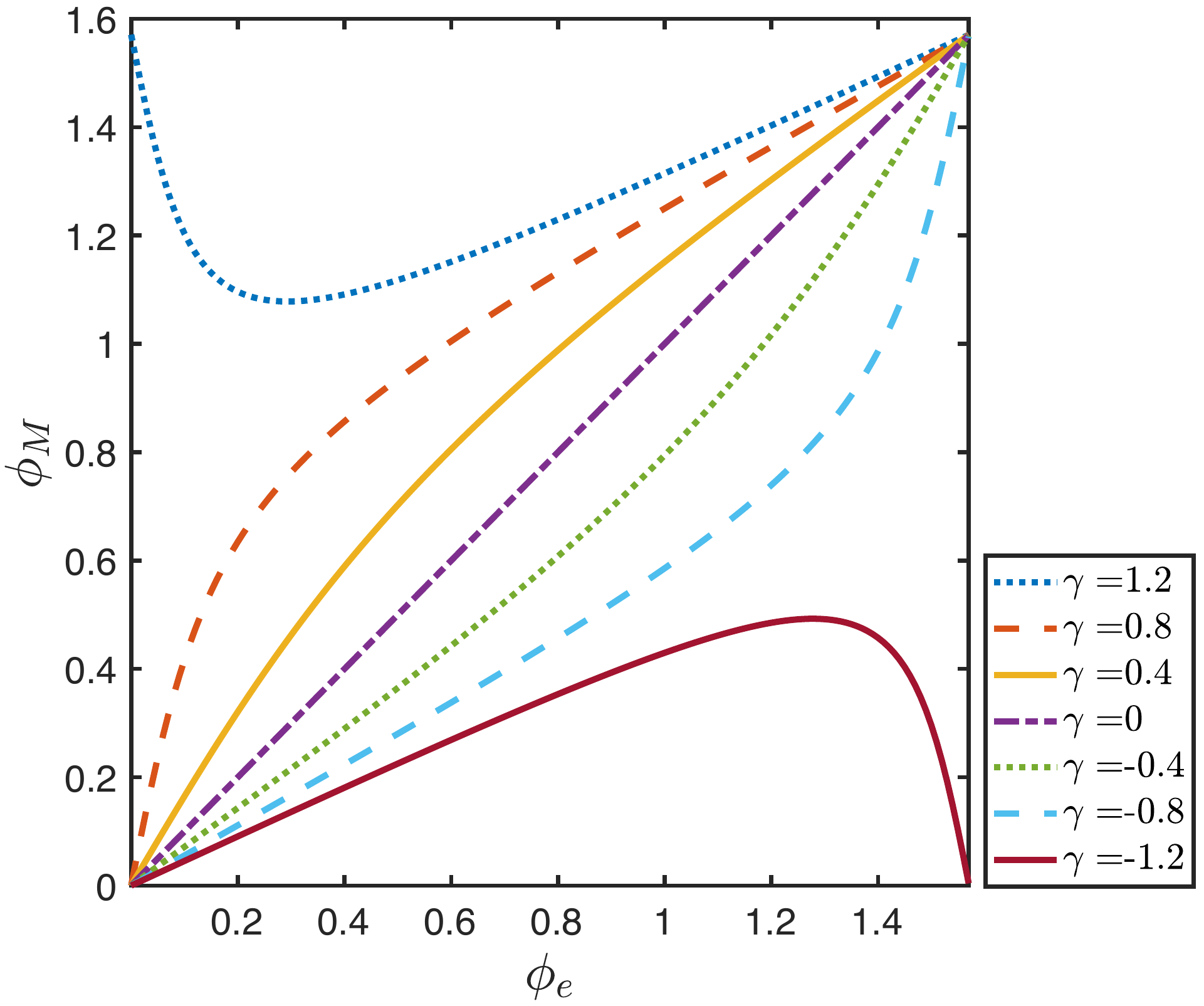}
\caption{Preferred direction of magnetization determined by $\phi_M$ as a function of the angle of the easy axis, $\phi_e$. Different curves illustrate different values of the parameter $\gamma$ from Eq.~(\ref{eq:gamma}).}
\label{fig:4}
\end{figure}

{\bf Without an external magnetic field.}
First, we consider the system without an external field ($H=0$)
\begin{equation}
\frac{\mathcal{E}}{\mathcal{V} K_u}=\sin^2(\phi_M-\phi_e)-\frac{\alpha \cos^2(\theta_P) }{K_u d} \sin^2(\phi_M).
\end{equation} 
For $\phi_e=l\pi/2$ ($l$ is any integer), the minimum of the energy is for either in-plane or out-of-plane direction of $\mathbf{M}$. For example, for $\phi_e=0$, we have 
\begin{equation}
\frac{\mathcal{E}}{K_u \mathcal{V}}=\sin^2(\phi_M)\left(1-\gamma \right).
\end{equation}
The defining dimensionless parameter here is 
\begin{equation}
\gamma=\frac{\alpha \cos^2(\theta_P)}{K_u d}.
\label{eq:gamma}
\end{equation}
Note that by changing the tilt of the molecules, we can change the direction of the easy axis of the magnet, assuming that $\alpha /(K_u d)>1$, see Fig.~\ref{fig:4}.   

For general values of $\phi_e$,
the equilibrium is reached when the derivative of the energy with respect to the magnetization direction is zero, $\partial \mathcal{E}/\partial \phi_M=0$, so that
\begin{equation}
\sin(2\phi_M-2\phi_e)-\gamma \sin(2\phi_M)=0.
\end{equation}
We see that the direction of the equilibrium direction of magnetization changes, which can explain the experimental observation of Ref.~\onlinecite{Meirzada2021}. For example, for $\alpha\to 0$ ($\gamma\to0$), we can write 
\begin{equation}
\phi_M\simeq \phi_e+ \gamma \sin(2\phi_e).
\end{equation}
For general values of $\gamma$, we solve the equation numerically, see Fig.~\ref{fig:4}. Note that the effect of small values of $\gamma$ is the most pronounced for $\phi_e\simeq \pi/4$.

{\bf With an external magnetic field.} The extended Stoner-Wohlfarth model provides insight into the modification of the coercive field caused by  the adsorbed molecules~\cite{Miwa_2020}. To illustrate this, let us assume 
an out-of-plane easy axis, $\phi_e = 0$,
and that the magnetic field is also perpendicular to the surface $\phi_H=0$. In this case, the organic molecular layer simply changes the value of the anisotropy parameter in the standard Stoner-Wohlfarth model, i.e., $K_u\to K_u(\alpha)$ where
\begin{equation}
K_u(\alpha)= K_u(\alpha=0)-\frac{\alpha \cos^2(\theta_P)}{d}.
\end{equation}

\begin{figure}
\includegraphics[scale=0.7]{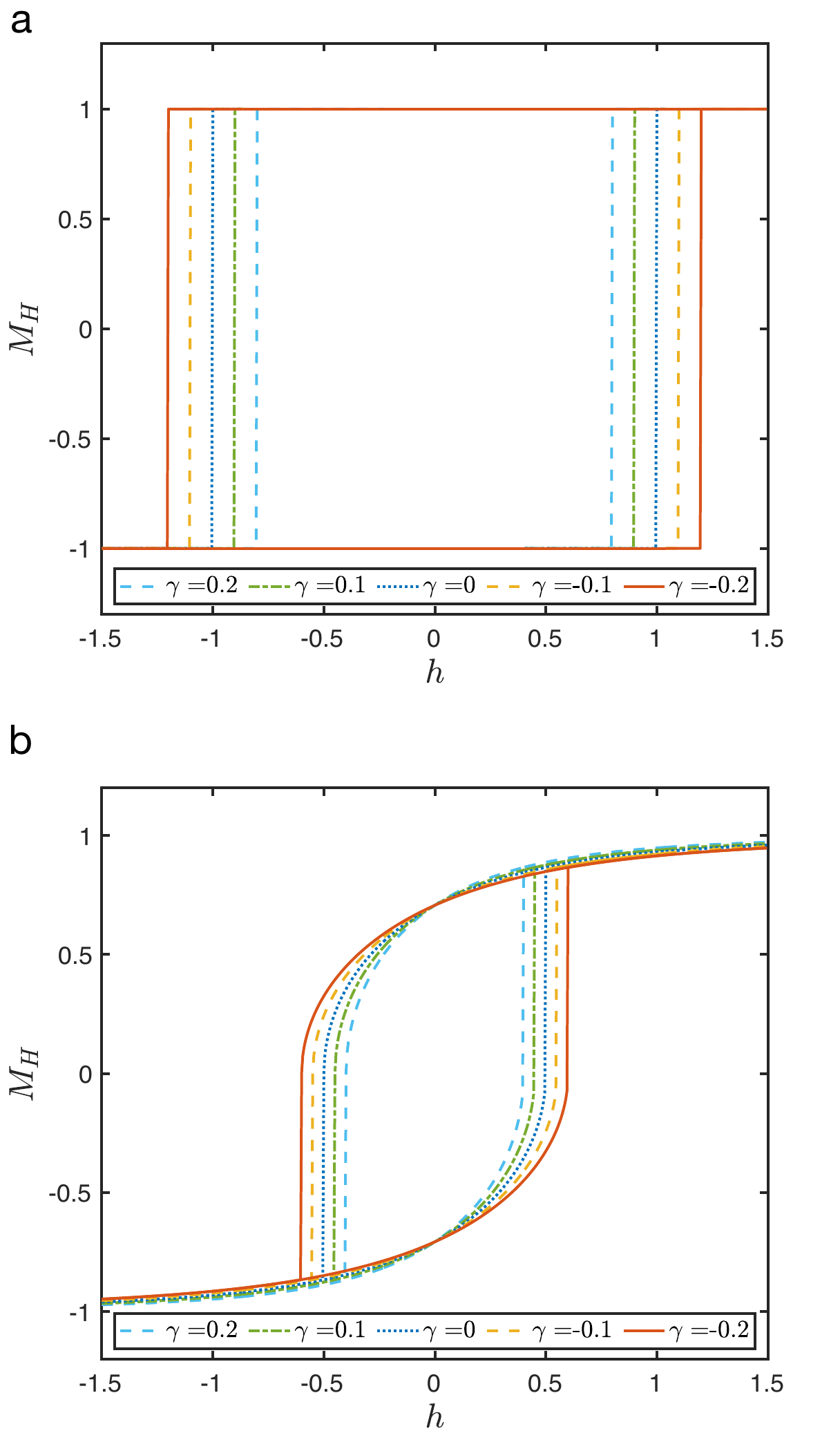}
\caption{Hysteresis loop in the ferromagnet, where $M_H = (\phi_M - \phi_H)$ is the magnetization projection on the magnetic field vector and $h = \mu_0 M_s H/2K_u$. a: Hysteresis for $\phi_H = 0$ and different values of the parameter $\gamma$. b: Same for $\phi_H = \pi/4 $.}
\label{fig:hyster}
\end{figure}

The change of $K_u$ affects the magnetic hysteresis, in particular, 
 the switching field (recall that $\phi_e=\phi_H=0$)
 \begin{equation}
 H_s(\alpha)=\frac{2 K_u(\alpha)}{\mu_0 M_s}.
 \end{equation}
The relative strength of the effect of the adsorbed molecules on the switching field is (see Fig.~\ref{fig:hyster}~a)
\begin{equation}
\frac{H_s(\alpha)}{H_s(0)}=1-\frac{\alpha \cos^2(\theta_P)}{d K_u(\alpha=0)}. 
\label{eq:switching_field}
\end{equation}
For general values of $\phi_e$, the Stoner-Wohlfarth model is not exactly solvable. Our numerical calculations show that the effect of the parameter $\gamma=\alpha\cos^2(\theta_P)/(d K_u(\alpha=0))$ is qualitatively similar to the case of $\phi_e=0$, see Fig.~\ref{fig:hyster}~b.

Using Eq.~(\ref{eq:switching_field}) and the experimental data~\cite{Miwa_2020}, we can have an-order-of-magnitude estimate for the value of $\alpha$, which is otherwise difficult to calculate because the strength of the SOC effect strongly depends on the electric field, whose accurate value is unknown. 

The substrate of thickness $d\simeq 0.7$ nm is made of iron for which\cite{Cullity2008} $K_u\simeq 3\times 10^{-4}$ eV/$\text{nm}^3$. 
Since the change in the coercive field
can be about 10\% (so that $\gamma\simeq 0.1$), we estimate $\alpha\simeq 2\times 10^{-5}$eV/nm$^2$.
Here, we have assumed that the molecules are perpendicular to the surface, $\theta_P=0$.  
Using the value of $\alpha$ and the Fermi energy, $\mathcal{E}_F\sim 11$eV, we can also estimate the strength of the Rashba SOC: $\alpha_R E_z\sim 10^{-4} \text{eV} \times$ nm -- a relatively small value\cite{manchon-2015}.
Note that if there is an electric field present without the molecules, i.e., $E_z\simeq E_z^0+\cos(\theta_P)E_z^1$, then we actually estimate $\alpha_R \sqrt{E_z^0 E_z^1}$ in this way.

\section{Conclusions and Outlook}
We have argued that the change of the surface electric field caused by adsorbed molecules can affect the magnetocrystalline anisotropy of a ferromagnetic metallic substrate even if molecules are achiral. 
To illustrate this claim,
we have formulated a toy model in which the molecules change the surface electric field, which is coupled to magnetization via a Rashba-like SOC. The strength of coupling is given by the
properties of the substrate and therefore no effect should be present for substrates with weak SOC  (cf.~Refs.~\onlinecite{Gersten2013,Liu2021,adhikari2022interplay}). 

Using the results of the toy model, we have introduced an extension of the Stoner-Wohlfarth framework, and calculated the effect of the molecules on the preferred direction of magnetization. In particular, we have shown that the effect of the molecules (for small values of $\gamma$) can be enhanced by working with a ferromagnet whose easy axis has $\phi_e\simeq \pi/4$, motivating experiments with such materials.
Finally, we have illustrated the effect of the molecules on the magnetic hysteresis, and estimated phenomenological parameters using available experimental data.  
Our findings show that isolating the effect of molecular chirality by changing magnetization of the surface is a daunting task, thus
motivating further experimental research 
of  chirality induced spin selectivity in current-free set-ups to use both enantiomers. 

Let us conclude by making a few remarks. First, note that according to our results the direction of magnetization couples to the dipolar moment of both chiral and achiral molecules, and therefore, our results cannot unravel microscopic origin of the CISS effect. Second, our work does not intend to 
discard the role of chirality in CISS experiments that explore the 
realignment of magnetization upon the adsorption of a chiral monolayer~\cite{BenDor2017}. On the
contrary, our work aims to motivate further theoretical studies that must contrast and compare
the roles of structural chirality and a magneto-electric coupling
driven by interface SOC. For instance, the term $\beta (\mathbf{P}\cdot\mathbf{n})^2(\mathbf{M}\cdot\mathbf{n})^2$ suggests that $\mathbf{M}$ can be either parallel or anti-parallel to $\mathbf{n}$. In other words, these two directions are energetically degenerate. Chirality then might be a key ingredient to break this degeneracy during the adsorption process leading to the results reported in Ref.~\onlinecite{BenDor2017}. The symmetry-breaking mechanism can be based upon an observation that the helical structure of the molecule is coupled to the angular momentum degree of freedom of the electron~\cite{Gersten2013,Liu2021,adhikari2022interplay}. One can speculate that this coupling in the presence of interface SOC can provide an energy barrier during molecular adsorption that depends on chirality and the direction of magnetization, and hence can potentially drive the system into only one of the available energy minima, i.e., with $\mathbf{M}$ either parallel or antiparallel to $\mathbf{n}$ depending on chirality.

Although the above mentioned frameworks\cite{Liu2021,adhikari2022interplay} deal with
time-dependent setups, they might pave the way for a natural extension of 
our model. Indeed, if we assume that the role of chirality is to filter the orbital angular momentum, then we can implement this filtering in SOC. 
Another possible extension of our work is to take into account a helical structure of the molecule, which adds to the 
electric field $\mathbf{E}$ some additional spatial
chirality.
%
%
To this end, one should study
more general forms of the electric field, e.g.,  $E_z=E^0_z+E^1_z\cos(\theta_P)+...$, including the terms independent of the tilting angle. Such terms can noticeably modify the overall dependence of observables on $\theta_P$. 
Furthermore, if we consider higher-order electric multipoles it should be possible to also study the effect of chirality. Finally, it is worth investigating the role of the electric field generated by the molecules on the CISS effect with currents, and the effect of other, in particular non-linear, types of SOC that can be present in the substrate.

\begin{acknowledgments}
We thank Zhanybek Alpichshev, Mohammad Reza Safari, Binghai Yan and Yossi Paltiel  for enlightning discussions. 

 M.L. acknowledges support by the European Research
Council (ERC) Starting Grant No.801770 (ANGULON).
A.~C. received funding from the European Union’s
Horizon Europe research and innovation program under the
Marie Skłodowska-Curie grant agreement No. 101062862 - NeqMolRot.
\end{acknowledgments}



\appendix

\section{Electric field of a single dipole}
\label{app:1}

\begin{figure}
\includegraphics[scale=0.5]{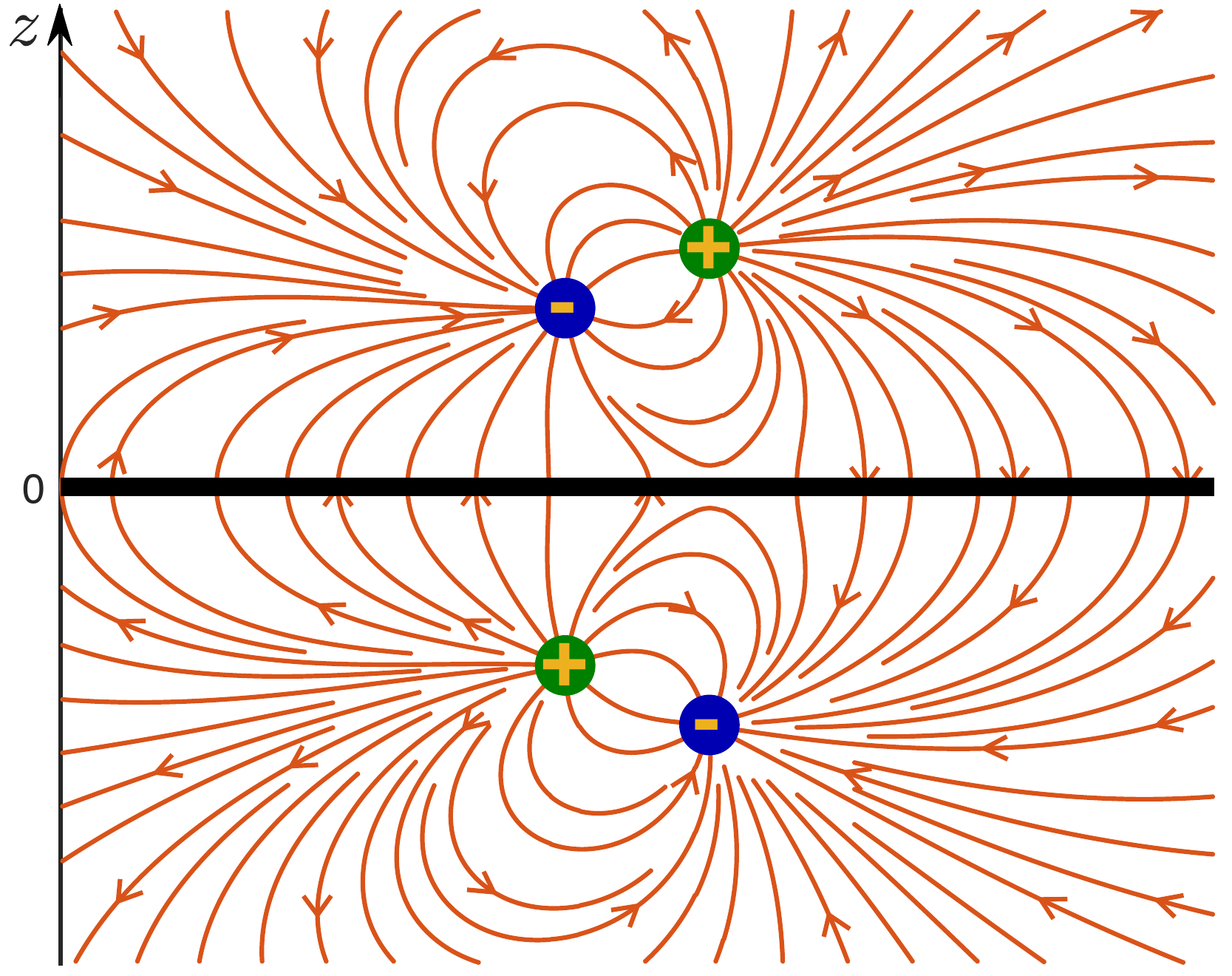}
\hfill
\caption{Electric field lines (the curves at $Z>0$) created by two charges (represented by the two balls at $Z>0$) above a metalic surface (the thick line at $Z=0$). The normal to the surface is along the $Z$-axis. The two charges placed at $Z<0$ are auxiliary mirror charges. The lines at $Z<0$ do not represent the physical electric field, and are shown only for illustrative purposes. }
\label{fig:app}
\end{figure}
Here, we discuss the electric field on the surface of a metal created by a single dipole. The geometry is shown in Fig.~\ref{fig:app}.
The corresponding electric field is (in cgs units):
\begin{equation}
E_z=\frac{2 q z_1}{r_1^3}-\frac{2 q z_2}{r_2^{3}},
\end{equation}
where $z_i$ is the position of the charge above the ferromagnet, and $r_i$ is the distance from the charge to the point on the surface, see Fig.~\ref{fig:app}.  
For simplicity, we assume a point dipole, $\mathbf{d}$, so that 
$q (z_1-z_2)=d_z$, and $E_z=2d_z/r_1^3$.
This expression shows that the electric field that enters our calculations indeed depends on the direction of the dipoles. 

We note that the realistic electric field generated by the molecules is unlikely to be that of a point dipole. It is also not homogeneous, in particular, because the molecules are extended entities. However, the qualitative properties of our main findings should be insensitive to these assumptions.

\bibliography{aipsamp}

\providecommand{\noopsort}[1]{}\providecommand{\singleletter}[1]{#1}%
\begin{thebibliography}{61}%
\makeatletter
\providecommand \@ifxundefined [1]{%
 \@ifx{#1\undefined}
}%
\providecommand \@ifnum [1]{%
 \ifnum #1\expandafter \@firstoftwo
 \else \expandafter \@secondoftwo
 \fi
}%
\providecommand \@ifx [1]{%
 \ifx #1\expandafter \@firstoftwo
 \else \expandafter \@secondoftwo
 \fi
}%
\providecommand \natexlab [1]{#1}%
\providecommand \enquote  [1]{``#1''}%
\providecommand \bibnamefont  [1]{#1}%
\providecommand \bibfnamefont [1]{#1}%
\providecommand \citenamefont [1]{#1}%
\providecommand \href@noop [0]{\@secondoftwo}%
\providecommand \href [0]{\begingroup \@sanitize@url \@href}%
\providecommand \@href[1]{\@@startlink{#1}\@@href}%
\providecommand \@@href[1]{\endgroup#1\@@endlink}%
\providecommand \@sanitize@url [0]{\catcode `\\12\catcode `\$12\catcode `\&12\catcode `\#12\catcode `\^12\catcode `\_12\catcode `\%12\relax}%
\providecommand \@@startlink[1]{}%
\providecommand \@@endlink[0]{}%
\providecommand \url  [0]{\begingroup\@sanitize@url \@url }%
\providecommand \@url [1]{\endgroup\@href {#1}{\urlprefix }}%
\providecommand \urlprefix  [0]{URL }%
\providecommand \Eprint [0]{\href }%
\providecommand \doibase [0]{http://dx.doi.org/}%
\providecommand \selectlanguage [0]{\@gobble}%
\providecommand \bibinfo  [0]{\@secondoftwo}%
\providecommand \bibfield  [0]{\@secondoftwo}%
\providecommand \translation [1]{[#1]}%
\providecommand \BibitemOpen [0]{}%
\providecommand \bibitemStop [0]{}%
\providecommand \bibitemNoStop [0]{.\EOS\space}%
\providecommand \EOS [0]{\spacefactor3000\relax}%
\providecommand \BibitemShut  [1]{\csname bibitem#1\endcsname}%
\let\auto@bib@innerbib\@empty
\bibitem [{\citenamefont {Naaman}, \citenamefont {Paltiel},\ and\ \citenamefont {Waldeck}(2019)}]{Naaman2019}%
  \BibitemOpen
  \bibfield  {author} {\bibinfo {author} {\bibfnamefont {R.}~\bibnamefont {Naaman}}, \bibinfo {author} {\bibfnamefont {Y.}~\bibnamefont {Paltiel}}, \ and\ \bibinfo {author} {\bibfnamefont {D.~H.}\ \bibnamefont {Waldeck}},\ }\bibfield  {title} {\enquote {\bibinfo {title} {Chiral molecules and the electron spin},}\ }\href {\doibase 10.1038/s41570-019-0087-1} {\bibfield  {journal} {\bibinfo  {journal} {Nature Reviews Chemistry}\ }\textbf {\bibinfo {volume} {3}},\ \bibinfo {pages} {250--260} (\bibinfo {year} {2019})}\BibitemShut {NoStop}%
\bibitem [{\citenamefont {Waldeck}, \citenamefont {Naaman},\ and\ \citenamefont {Paltiel}(2021)}]{Waldeck2021}%
  \BibitemOpen
  \bibfield  {author} {\bibinfo {author} {\bibfnamefont {D.~H.}\ \bibnamefont {Waldeck}}, \bibinfo {author} {\bibfnamefont {R.}~\bibnamefont {Naaman}}, \ and\ \bibinfo {author} {\bibfnamefont {Y.}~\bibnamefont {Paltiel}},\ }\bibfield  {title} {\enquote {\bibinfo {title} {{The spin selectivity effect in chiral materials}},}\ }\href {\doibase 10.1063/5.0049150} {\bibfield  {journal} {\bibinfo  {journal} {APL Materials}\ }\textbf {\bibinfo {volume} {9}} (\bibinfo {year} {2021}),\ 10.1063/5.0049150}\BibitemShut {NoStop}%
\bibitem [{\citenamefont {Naaman}, \citenamefont {Paltiel},\ and\ \citenamefont {Waldeck}(2022)}]{Naaman2022}%
  \BibitemOpen
  \bibfield  {author} {\bibinfo {author} {\bibfnamefont {R.}~\bibnamefont {Naaman}}, \bibinfo {author} {\bibfnamefont {Y.}~\bibnamefont {Paltiel}}, \ and\ \bibinfo {author} {\bibfnamefont {D.~H.}\ \bibnamefont {Waldeck}},\ }\bibfield  {title} {\enquote {\bibinfo {title} {Chiral induced spin selectivity and its implications for biological functions},}\ }\href {\doibase 10.1146/annurev-biophys-083021-070400} {\bibfield  {journal} {\bibinfo  {journal} {Annual Review of Biophysics}\ }\textbf {\bibinfo {volume} {51}},\ \bibinfo {pages} {99--114} (\bibinfo {year} {2022})}\BibitemShut {NoStop}%
\bibitem [{\citenamefont {Xu}\ and\ \citenamefont {Mi}(2023)}]{Xu2023}%
  \BibitemOpen
  \bibfield  {author} {\bibinfo {author} {\bibfnamefont {Y.}~\bibnamefont {Xu}}\ and\ \bibinfo {author} {\bibfnamefont {W.}~\bibnamefont {Mi}},\ }\bibfield  {title} {\enquote {\bibinfo {title} {Chiral-induced spin selectivity in biomolecules{,} hybrid organic–inorganic perovskites and inorganic materials: a comprehensive review on recent progress},}\ }\href {\doibase 10.1039/D3MH00024A} {\bibfield  {journal} {\bibinfo  {journal} {Mater. Horiz.}\ ,\ \bibinfo {pages} {--}} (\bibinfo {year} {2023})}\BibitemShut {NoStop}%
\bibitem [{\citenamefont {Ray}\ \emph {et~al.}(1999)\citenamefont {Ray}, \citenamefont {Ananthavel}, \citenamefont {Waldeck},\ and\ \citenamefont {Naaman}}]{Ray1999}%
  \BibitemOpen
  \bibfield  {author} {\bibinfo {author} {\bibfnamefont {K.}~\bibnamefont {Ray}}, \bibinfo {author} {\bibfnamefont {S.~P.}\ \bibnamefont {Ananthavel}}, \bibinfo {author} {\bibfnamefont {D.~H.}\ \bibnamefont {Waldeck}}, \ and\ \bibinfo {author} {\bibfnamefont {R.}~\bibnamefont {Naaman}},\ }\bibfield  {title} {\enquote {\bibinfo {title} {Asymmetric scattering of polarized electrons by organized organic films of chiral molecules},}\ }\href {\doibase 10.1126/science.283.5403.814} {\bibfield  {journal} {\bibinfo  {journal} {Science}\ }\textbf {\bibinfo {volume} {283}},\ \bibinfo {pages} {814--816} (\bibinfo {year} {1999})}\BibitemShut {NoStop}%
\bibitem [{\citenamefont {G\"{o}hler}\ \emph {et~al.}(2011)\citenamefont {G\"{o}hler}, \citenamefont {Hamelbeck}, \citenamefont {Markus}, \citenamefont {Kettner}, \citenamefont {Hanne}, \citenamefont {Vager}, \citenamefont {Naaman},\ and\ \citenamefont {Zacharias}}]{Ghler2011}%
  \BibitemOpen
  \bibfield  {author} {\bibinfo {author} {\bibfnamefont {B.}~\bibnamefont {G\"{o}hler}}, \bibinfo {author} {\bibfnamefont {V.}~\bibnamefont {Hamelbeck}}, \bibinfo {author} {\bibfnamefont {T.~Z.}\ \bibnamefont {Markus}}, \bibinfo {author} {\bibfnamefont {M.}~\bibnamefont {Kettner}}, \bibinfo {author} {\bibfnamefont {G.~F.}\ \bibnamefont {Hanne}}, \bibinfo {author} {\bibfnamefont {Z.}~\bibnamefont {Vager}}, \bibinfo {author} {\bibfnamefont {R.}~\bibnamefont {Naaman}}, \ and\ \bibinfo {author} {\bibfnamefont {H.}~\bibnamefont {Zacharias}},\ }\bibfield  {title} {\enquote {\bibinfo {title} {Spin selectivity in electron transmission through self-assembled monolayers of double-stranded {DNA}},}\ }\href {\doibase 10.1126/science.1199339} {\bibfield  {journal} {\bibinfo  {journal} {Science}\ }\textbf {\bibinfo {volume} {331}},\ \bibinfo {pages} {894--897} (\bibinfo {year} {2011})}\BibitemShut {NoStop}%
\bibitem [{\citenamefont {Xie}\ \emph {et~al.}(2011)\citenamefont {Xie}, \citenamefont {Markus}, \citenamefont {Cohen}, \citenamefont {Vager}, \citenamefont {Gutierrez},\ and\ \citenamefont {Naaman}}]{Xie2011}%
  \BibitemOpen
  \bibfield  {author} {\bibinfo {author} {\bibfnamefont {Z.}~\bibnamefont {Xie}}, \bibinfo {author} {\bibfnamefont {T.~Z.}\ \bibnamefont {Markus}}, \bibinfo {author} {\bibfnamefont {S.~R.}\ \bibnamefont {Cohen}}, \bibinfo {author} {\bibfnamefont {Z.}~\bibnamefont {Vager}}, \bibinfo {author} {\bibfnamefont {R.}~\bibnamefont {Gutierrez}}, \ and\ \bibinfo {author} {\bibfnamefont {R.}~\bibnamefont {Naaman}},\ }\bibfield  {title} {\enquote {\bibinfo {title} {Spin specific electron conduction through {DNA} oligomers},}\ }\href {\doibase 10.1021/nl2021637} {\bibfield  {journal} {\bibinfo  {journal} {Nano Letters}\ }\textbf {\bibinfo {volume} {11}},\ \bibinfo {pages} {4652--4655} (\bibinfo {year} {2011})}\BibitemShut {NoStop}%
\bibitem [{\citenamefont {Zhang}\ \emph {et~al.}(2018)\citenamefont {Zhang}, \citenamefont {Banerjee-Ghosh}, \citenamefont {Tassinari},\ and\ \citenamefont {Naaman}}]{Zhang2018}%
  \BibitemOpen
  \bibfield  {author} {\bibinfo {author} {\bibfnamefont {W.}~\bibnamefont {Zhang}}, \bibinfo {author} {\bibfnamefont {K.}~\bibnamefont {Banerjee-Ghosh}}, \bibinfo {author} {\bibfnamefont {F.}~\bibnamefont {Tassinari}}, \ and\ \bibinfo {author} {\bibfnamefont {R.}~\bibnamefont {Naaman}},\ }\bibfield  {title} {\enquote {\bibinfo {title} {Enhanced electrochemical water splitting with chiral molecule-coated fe3o4 nanoparticles},}\ }\href {\doibase 10.1021/acsenergylett.8b01454} {\bibfield  {journal} {\bibinfo  {journal} {ACS Energy Letters}\ }\textbf {\bibinfo {volume} {3}},\ \bibinfo {pages} {2308--2313} (\bibinfo {year} {2018})}\BibitemShut {NoStop}%
\bibitem [{\citenamefont {Metzger}\ \emph {et~al.}(2020)\citenamefont {Metzger}, \citenamefont {Mishra}, \citenamefont {Bloom}, \citenamefont {Goren}, \citenamefont {Neubauer}, \citenamefont {Shmul}, \citenamefont {Wei}, \citenamefont {Yochelis}, \citenamefont {Tassinari}, \citenamefont {Fontanesi}, \citenamefont {Waldeck}, \citenamefont {Paltiel},\ and\ \citenamefont {Naaman}}]{Metzger2020}%
  \BibitemOpen
  \bibfield  {author} {\bibinfo {author} {\bibfnamefont {T.~S.}\ \bibnamefont {Metzger}}, \bibinfo {author} {\bibfnamefont {S.}~\bibnamefont {Mishra}}, \bibinfo {author} {\bibfnamefont {B.~P.}\ \bibnamefont {Bloom}}, \bibinfo {author} {\bibfnamefont {N.}~\bibnamefont {Goren}}, \bibinfo {author} {\bibfnamefont {A.}~\bibnamefont {Neubauer}}, \bibinfo {author} {\bibfnamefont {G.}~\bibnamefont {Shmul}}, \bibinfo {author} {\bibfnamefont {J.}~\bibnamefont {Wei}}, \bibinfo {author} {\bibfnamefont {S.}~\bibnamefont {Yochelis}}, \bibinfo {author} {\bibfnamefont {F.}~\bibnamefont {Tassinari}}, \bibinfo {author} {\bibfnamefont {C.}~\bibnamefont {Fontanesi}}, \bibinfo {author} {\bibfnamefont {D.~H.}\ \bibnamefont {Waldeck}}, \bibinfo {author} {\bibfnamefont {Y.}~\bibnamefont {Paltiel}}, \ and\ \bibinfo {author} {\bibfnamefont {R.}~\bibnamefont {Naaman}},\ }\bibfield  {title} {\enquote {\bibinfo {title} {The electron spin as a chiral reagent},}\ }\href {\doibase https://doi.org/10.1002/anie.201911400} {\bibfield  {journal} {\bibinfo  {journal} {Angewandte Chemie International Edition}\ }\textbf {\bibinfo {volume} {59}},\ \bibinfo {pages} {1653--1658} (\bibinfo {year} {2020})}\BibitemShut {NoStop}%
\bibitem [{\citenamefont {Banerjee-Ghosh}\ \emph {et~al.}(2018)\citenamefont {Banerjee-Ghosh}, \citenamefont {Dor}, \citenamefont {Tassinari}, \citenamefont {Capua}, \citenamefont {Yochelis}, \citenamefont {Capua}, \citenamefont {Yang}, \citenamefont {Parkin}, \citenamefont {Sarkar}, \citenamefont {Kronik}, \citenamefont {Baczewski}, \citenamefont {Naaman},\ and\ \citenamefont {Paltiel}}]{BanerjeeGhosh2018}%
  \BibitemOpen
  \bibfield  {author} {\bibinfo {author} {\bibfnamefont {K.}~\bibnamefont {Banerjee-Ghosh}}, \bibinfo {author} {\bibfnamefont {O.~B.}\ \bibnamefont {Dor}}, \bibinfo {author} {\bibfnamefont {F.}~\bibnamefont {Tassinari}}, \bibinfo {author} {\bibfnamefont {E.}~\bibnamefont {Capua}}, \bibinfo {author} {\bibfnamefont {S.}~\bibnamefont {Yochelis}}, \bibinfo {author} {\bibfnamefont {A.}~\bibnamefont {Capua}}, \bibinfo {author} {\bibfnamefont {S.-H.}\ \bibnamefont {Yang}}, \bibinfo {author} {\bibfnamefont {S.~S.~P.}\ \bibnamefont {Parkin}}, \bibinfo {author} {\bibfnamefont {S.}~\bibnamefont {Sarkar}}, \bibinfo {author} {\bibfnamefont {L.}~\bibnamefont {Kronik}}, \bibinfo {author} {\bibfnamefont {L.~T.}\ \bibnamefont {Baczewski}}, \bibinfo {author} {\bibfnamefont {R.}~\bibnamefont {Naaman}}, \ and\ \bibinfo {author} {\bibfnamefont {Y.}~\bibnamefont {Paltiel}},\ }\bibfield  {title} {\enquote {\bibinfo {title} {Separation of enantiomers by their enantiospecific interaction with achiral magnetic substrates},}\ }\href {\doibase 10.1126/science.aar4265} {\bibfield  {journal} {\bibinfo  {journal} {Science}\ }\textbf {\bibinfo {volume} {360}},\ \bibinfo {pages} {1331--1334} (\bibinfo {year} {2018})}\BibitemShut {NoStop}%
\bibitem [{\citenamefont {Safari}\ \emph {et~al.}(2022)\citenamefont {Safari}, \citenamefont {Matthes}, \citenamefont {Ernst}, \citenamefont {Bürgler},\ and\ \citenamefont {Schneider}}]{safari2022enantiospecific}%
  \BibitemOpen
  \bibfield  {author} {\bibinfo {author} {\bibfnamefont {M.~R.}\ \bibnamefont {Safari}}, \bibinfo {author} {\bibfnamefont {F.}~\bibnamefont {Matthes}}, \bibinfo {author} {\bibfnamefont {K.-H.}\ \bibnamefont {Ernst}}, \bibinfo {author} {\bibfnamefont {D.~E.}\ \bibnamefont {Bürgler}}, \ and\ \bibinfo {author} {\bibfnamefont {C.~M.}\ \bibnamefont {Schneider}},\ }\href@noop {} {\enquote {\bibinfo {title} {Enantiospecific adsorption on a ferromagnetic surface at the single-molecule scale},}\ } (\bibinfo {year} {2022}),\ \Eprint {http://arxiv.org/abs/2211.12976} {arXiv:2211.12976 [cond-mat.mtrl-sci]} \BibitemShut {NoStop}%
\bibitem [{\citenamefont {Santra}\ \emph {et~al.}(2023)\citenamefont {Santra}, \citenamefont {Lu}, \citenamefont {Waldeck},\ and\ \citenamefont {Naaman}}]{Santra2023}%
  \BibitemOpen
  \bibfield  {author} {\bibinfo {author} {\bibfnamefont {K.}~\bibnamefont {Santra}}, \bibinfo {author} {\bibfnamefont {Y.}~\bibnamefont {Lu}}, \bibinfo {author} {\bibfnamefont {D.~H.}\ \bibnamefont {Waldeck}}, \ and\ \bibinfo {author} {\bibfnamefont {R.}~\bibnamefont {Naaman}},\ }\bibfield  {title} {\enquote {\bibinfo {title} {Spin selectivity damage dependence of adsorption of dsdna on ferromagnets},}\ }\href {\doibase 10.1021/acs.jpcb.2c08820} {\bibfield  {journal} {\bibinfo  {journal} {The Journal of Physical Chemistry B}\ }\textbf {\bibinfo {volume} {127}},\ \bibinfo {pages} {2344--2350} (\bibinfo {year} {2023})},\ \bibinfo {note} {pMID: 36888909}\BibitemShut {NoStop}%
\bibitem [{\citenamefont {Ghosh}\ \emph {et~al.}(2019)\citenamefont {Ghosh}, \citenamefont {Zhang}, \citenamefont {Tassinari}, \citenamefont {Mastai}, \citenamefont {Lidor-Shalev}, \citenamefont {Naaman}, \citenamefont {Möllers}, \citenamefont {Nürenberg}, \citenamefont {Zacharias}, \citenamefont {Wei}, \citenamefont {Wierzbinski},\ and\ \citenamefont {Waldeck}}]{Ghosh2019}%
  \BibitemOpen
  \bibfield  {author} {\bibinfo {author} {\bibfnamefont {K.~B.}\ \bibnamefont {Ghosh}}, \bibinfo {author} {\bibfnamefont {W.}~\bibnamefont {Zhang}}, \bibinfo {author} {\bibfnamefont {F.}~\bibnamefont {Tassinari}}, \bibinfo {author} {\bibfnamefont {Y.}~\bibnamefont {Mastai}}, \bibinfo {author} {\bibfnamefont {O.}~\bibnamefont {Lidor-Shalev}}, \bibinfo {author} {\bibfnamefont {R.}~\bibnamefont {Naaman}}, \bibinfo {author} {\bibfnamefont {P.}~\bibnamefont {Möllers}}, \bibinfo {author} {\bibfnamefont {D.}~\bibnamefont {Nürenberg}}, \bibinfo {author} {\bibfnamefont {H.}~\bibnamefont {Zacharias}}, \bibinfo {author} {\bibfnamefont {J.}~\bibnamefont {Wei}}, \bibinfo {author} {\bibfnamefont {E.}~\bibnamefont {Wierzbinski}}, \ and\ \bibinfo {author} {\bibfnamefont {D.~H.}\ \bibnamefont {Waldeck}},\ }\bibfield  {title} {\enquote {\bibinfo {title} {Controlling chemical selectivity in electrocatalysis with chiral cuo-coated electrodes},}\ }\href {\doibase 10.1021/acs.jpcc.8b12027} {\bibfield  {journal} {\bibinfo  {journal} {The Journal of Physical Chemistry C}\ }\textbf {\bibinfo {volume} {123}},\ \bibinfo {pages} {3024--3031} (\bibinfo {year} {2019})}\BibitemShut {NoStop}%
\bibitem [{\citenamefont {Inui}\ \emph {et~al.}(2020)\citenamefont {Inui}, \citenamefont {Aoki}, \citenamefont {Nishiue}, \citenamefont {Shiota}, \citenamefont {Kousaka}, \citenamefont {Shishido}, \citenamefont {Hirobe}, \citenamefont {Suda}, \citenamefont {Ohe}, \citenamefont {Kishine}, \citenamefont {Yamamoto},\ and\ \citenamefont {Togawa}}]{Inui2020}%
  \BibitemOpen
  \bibfield  {author} {\bibinfo {author} {\bibfnamefont {A.}~\bibnamefont {Inui}}, \bibinfo {author} {\bibfnamefont {R.}~\bibnamefont {Aoki}}, \bibinfo {author} {\bibfnamefont {Y.}~\bibnamefont {Nishiue}}, \bibinfo {author} {\bibfnamefont {K.}~\bibnamefont {Shiota}}, \bibinfo {author} {\bibfnamefont {Y.}~\bibnamefont {Kousaka}}, \bibinfo {author} {\bibfnamefont {H.}~\bibnamefont {Shishido}}, \bibinfo {author} {\bibfnamefont {D.}~\bibnamefont {Hirobe}}, \bibinfo {author} {\bibfnamefont {M.}~\bibnamefont {Suda}}, \bibinfo {author} {\bibfnamefont {J.-i.}\ \bibnamefont {Ohe}}, \bibinfo {author} {\bibfnamefont {J.-i.}\ \bibnamefont {Kishine}}, \bibinfo {author} {\bibfnamefont {H.~M.}\ \bibnamefont {Yamamoto}}, \ and\ \bibinfo {author} {\bibfnamefont {Y.}~\bibnamefont {Togawa}},\ }\bibfield  {title} {\enquote {\bibinfo {title} {{Chirality-Induced Spin-Polarized State of a Chiral Crystal ${\mathrm{CrNb}}_{3}{\mathrm{S}}_{6}$}},}\ }\href {\doibase 10.1103/PhysRevLett.124.166602} {\bibfield  {journal} {\bibinfo  {journal} {Phys. Rev. Lett.}\ }\textbf {\bibinfo {volume} {124}},\ \bibinfo {pages} {166602} (\bibinfo {year} {2020})}\BibitemShut {NoStop}%
\bibitem [{\citenamefont {Evers}\ \emph {et~al.}(2022)\citenamefont {Evers}, \citenamefont {Aharony}, \citenamefont {Bar-Gill}, \citenamefont {Entin-Wohlman}, \citenamefont {Hedegård}, \citenamefont {Hod}, \citenamefont {Jelinek}, \citenamefont {Kamieniarz}, \citenamefont {Lemeshko}, \citenamefont {Michaeli}, \citenamefont {Mujica}, \citenamefont {Naaman}, \citenamefont {Paltiel}, \citenamefont {Refaely-Abramson}, \citenamefont {Tal}, \citenamefont {Thijssen}, \citenamefont {Thoss}, \citenamefont {van Ruitenbeek}, \citenamefont {Venkataraman}, \citenamefont {Waldeck}, \citenamefont {Yan},\ and\ \citenamefont {Kronik}}]{Evers2022}%
  \BibitemOpen
  \bibfield  {author} {\bibinfo {author} {\bibfnamefont {F.}~\bibnamefont {Evers}}, \bibinfo {author} {\bibfnamefont {A.}~\bibnamefont {Aharony}}, \bibinfo {author} {\bibfnamefont {N.}~\bibnamefont {Bar-Gill}}, \bibinfo {author} {\bibfnamefont {O.}~\bibnamefont {Entin-Wohlman}}, \bibinfo {author} {\bibfnamefont {P.}~\bibnamefont {Hedegård}}, \bibinfo {author} {\bibfnamefont {O.}~\bibnamefont {Hod}}, \bibinfo {author} {\bibfnamefont {P.}~\bibnamefont {Jelinek}}, \bibinfo {author} {\bibfnamefont {G.}~\bibnamefont {Kamieniarz}}, \bibinfo {author} {\bibfnamefont {M.}~\bibnamefont {Lemeshko}}, \bibinfo {author} {\bibfnamefont {K.}~\bibnamefont {Michaeli}}, \bibinfo {author} {\bibfnamefont {V.}~\bibnamefont {Mujica}}, \bibinfo {author} {\bibfnamefont {R.}~\bibnamefont {Naaman}}, \bibinfo {author} {\bibfnamefont {Y.}~\bibnamefont {Paltiel}}, \bibinfo {author} {\bibfnamefont {S.}~\bibnamefont {Refaely-Abramson}}, \bibinfo {author} {\bibfnamefont {O.}~\bibnamefont {Tal}}, \bibinfo {author} {\bibfnamefont {J.}~\bibnamefont {Thijssen}}, \bibinfo {author} {\bibfnamefont {M.}~\bibnamefont {Thoss}}, \bibinfo {author} {\bibfnamefont {J.~M.}\ \bibnamefont {van Ruitenbeek}}, \bibinfo {author} {\bibfnamefont {L.}~\bibnamefont {Venkataraman}}, \bibinfo {author} {\bibfnamefont {D.~H.}\ \bibnamefont {Waldeck}}, \bibinfo {author} {\bibfnamefont {B.}~\bibnamefont {Yan}}, \ and\ \bibinfo {author} {\bibfnamefont {L.}~\bibnamefont {Kronik}},\ }\bibfield  {title} {\enquote {\bibinfo {title} {Theory of chirality induced spin selectivity: Progress and challenges},}\ }\href {\doibase https://doi.org/10.1002/adma.202106629} {\bibfield  {journal} {\bibinfo  {journal} {Advanced Materials}\ }\textbf {\bibinfo {volume} {34}},\ \bibinfo {pages} {2106629} (\bibinfo {year} {2022})}\BibitemShut {NoStop}%
\bibitem [{\citenamefont {Fransson}(2022)}]{Fransson2022}%
  \BibitemOpen
  \bibfield  {author} {\bibinfo {author} {\bibfnamefont {J.}~\bibnamefont {Fransson}},\ }\bibfield  {title} {\enquote {\bibinfo {title} {The chiral induced spin selectivity effect what it is, what it is not, and why it matters},}\ }\href {\doibase https://doi.org/10.1002/ijch.202200046} {\bibfield  {journal} {\bibinfo  {journal} {Israel Journal of Chemistry}\ }\textbf {\bibinfo {volume} {62}},\ \bibinfo {pages} {e202200046} (\bibinfo {year} {2022})}\BibitemShut {NoStop}%
\bibitem [{\citenamefont {Sukenik}\ \emph {et~al.}(2020)\citenamefont {Sukenik}, \citenamefont {Tassinari}, \citenamefont {Yochelis}, \citenamefont {Millo}, \citenamefont {Baczewski},\ and\ \citenamefont {Paltiel}}]{Sukenik2020}%
  \BibitemOpen
  \bibfield  {author} {\bibinfo {author} {\bibfnamefont {N.}~\bibnamefont {Sukenik}}, \bibinfo {author} {\bibfnamefont {F.}~\bibnamefont {Tassinari}}, \bibinfo {author} {\bibfnamefont {S.}~\bibnamefont {Yochelis}}, \bibinfo {author} {\bibfnamefont {O.}~\bibnamefont {Millo}}, \bibinfo {author} {\bibfnamefont {L.~T.}\ \bibnamefont {Baczewski}}, \ and\ \bibinfo {author} {\bibfnamefont {Y.}~\bibnamefont {Paltiel}},\ }\bibfield  {title} {\enquote {\bibinfo {title} {Correlation between ferromagnetic layer easy axis and the tilt angle of self assembled chiral molecules},}\ }\href {\doibase 10.3390/molecules25246036} {\bibfield  {journal} {\bibinfo  {journal} {Molecules}\ }\textbf {\bibinfo {volume} {25}} (\bibinfo {year} {2020}),\ 10.3390/molecules25246036}\BibitemShut {NoStop}%
\bibitem [{\citenamefont {Miwa}\ \emph {et~al.}(2020)\citenamefont {Miwa}, \citenamefont {Kondou}, \citenamefont {Sakamoto}, \citenamefont {Nihonyanagi}, \citenamefont {Araoka}, \citenamefont {Otani},\ and\ \citenamefont {Miyajima}}]{Miwa_2020}%
  \BibitemOpen
  \bibfield  {author} {\bibinfo {author} {\bibfnamefont {S.}~\bibnamefont {Miwa}}, \bibinfo {author} {\bibfnamefont {K.}~\bibnamefont {Kondou}}, \bibinfo {author} {\bibfnamefont {S.}~\bibnamefont {Sakamoto}}, \bibinfo {author} {\bibfnamefont {A.}~\bibnamefont {Nihonyanagi}}, \bibinfo {author} {\bibfnamefont {F.}~\bibnamefont {Araoka}}, \bibinfo {author} {\bibfnamefont {Y.}~\bibnamefont {Otani}}, \ and\ \bibinfo {author} {\bibfnamefont {D.}~\bibnamefont {Miyajima}},\ }\bibfield  {title} {\enquote {\bibinfo {title} {Chirality-induced effective magnetic field in a phthalocyanine molecule},}\ }\href {\doibase 10.35848/1882-0786/abbf67} {\bibfield  {journal} {\bibinfo  {journal} {Applied Physics Express}\ }\textbf {\bibinfo {volume} {13}},\ \bibinfo {pages} {113001} (\bibinfo {year} {2020})}\BibitemShut {NoStop}%
\bibitem [{\citenamefont {Meirzada}\ \emph {et~al.}(2021)\citenamefont {Meirzada}, \citenamefont {Sukenik}, \citenamefont {Haim}, \citenamefont {Yochelis}, \citenamefont {Baczewski}, \citenamefont {Paltiel},\ and\ \citenamefont {Bar-Gill}}]{Meirzada2021}%
  \BibitemOpen
  \bibfield  {author} {\bibinfo {author} {\bibfnamefont {I.}~\bibnamefont {Meirzada}}, \bibinfo {author} {\bibfnamefont {N.}~\bibnamefont {Sukenik}}, \bibinfo {author} {\bibfnamefont {G.}~\bibnamefont {Haim}}, \bibinfo {author} {\bibfnamefont {S.}~\bibnamefont {Yochelis}}, \bibinfo {author} {\bibfnamefont {L.~T.}\ \bibnamefont {Baczewski}}, \bibinfo {author} {\bibfnamefont {Y.}~\bibnamefont {Paltiel}}, \ and\ \bibinfo {author} {\bibfnamefont {N.}~\bibnamefont {Bar-Gill}},\ }\bibfield  {title} {\enquote {\bibinfo {title} {Long-time-scale magnetization ordering induced by an adsorbed chiral monolayer on ferromagnets},}\ }\href {\doibase 10.1021/acsnano.1c00455} {\bibfield  {journal} {\bibinfo  {journal} {ACS Nano}\ }\textbf {\bibinfo {volume} {15}},\ \bibinfo {pages} {5574--5579} (\bibinfo {year} {2021})},\ \bibinfo {note} {pMID: 33591720}\BibitemShut {NoStop}%
\bibitem [{\citenamefont {Alpern}\ \emph {et~al.}(2021)\citenamefont {Alpern}, \citenamefont {Amundsen}, \citenamefont {Hartmann}, \citenamefont {Sukenik}, \citenamefont {Spuri}, \citenamefont {Yochelis}, \citenamefont {Prokscha}, \citenamefont {Gutkin}, \citenamefont {Anahory}, \citenamefont {Scheer}, \citenamefont {Linder}, \citenamefont {Salman}, \citenamefont {Millo}, \citenamefont {Paltiel},\ and\ \citenamefont {Di~Bernardo}}]{Alpern2021}%
  \BibitemOpen
  \bibfield  {author} {\bibinfo {author} {\bibfnamefont {H.}~\bibnamefont {Alpern}}, \bibinfo {author} {\bibfnamefont {M.}~\bibnamefont {Amundsen}}, \bibinfo {author} {\bibfnamefont {R.}~\bibnamefont {Hartmann}}, \bibinfo {author} {\bibfnamefont {N.}~\bibnamefont {Sukenik}}, \bibinfo {author} {\bibfnamefont {A.}~\bibnamefont {Spuri}}, \bibinfo {author} {\bibfnamefont {S.}~\bibnamefont {Yochelis}}, \bibinfo {author} {\bibfnamefont {T.}~\bibnamefont {Prokscha}}, \bibinfo {author} {\bibfnamefont {V.}~\bibnamefont {Gutkin}}, \bibinfo {author} {\bibfnamefont {Y.}~\bibnamefont {Anahory}}, \bibinfo {author} {\bibfnamefont {E.}~\bibnamefont {Scheer}}, \bibinfo {author} {\bibfnamefont {J.}~\bibnamefont {Linder}}, \bibinfo {author} {\bibfnamefont {Z.}~\bibnamefont {Salman}}, \bibinfo {author} {\bibfnamefont {O.}~\bibnamefont {Millo}}, \bibinfo {author} {\bibfnamefont {Y.}~\bibnamefont {Paltiel}}, \ and\ \bibinfo {author} {\bibfnamefont {A.}~\bibnamefont {Di~Bernardo}},\ }\bibfield  {title} {\enquote {\bibinfo {title} {Unconventional meissner screening induced by chiral molecules in a conventional superconductor},}\ }\href {\doibase 10.1103/PhysRevMaterials.5.114801} {\bibfield  {journal} {\bibinfo  {journal} {Phys. Rev. Mater.}\ }\textbf {\bibinfo {volume} {5}},\ \bibinfo {pages} {114801} (\bibinfo {year} {2021})}\BibitemShut {NoStop}%
\bibitem [{\citenamefont {Gutierrez}\ \emph {et~al.}(2012)\citenamefont {Gutierrez}, \citenamefont {D\'{\i}az}, \citenamefont {Naaman},\ and\ \citenamefont {Cuniberti}}]{Gutierrez2012}%
  \BibitemOpen
  \bibfield  {author} {\bibinfo {author} {\bibfnamefont {R.}~\bibnamefont {Gutierrez}}, \bibinfo {author} {\bibfnamefont {E.}~\bibnamefont {D\'{\i}az}}, \bibinfo {author} {\bibfnamefont {R.}~\bibnamefont {Naaman}}, \ and\ \bibinfo {author} {\bibfnamefont {G.}~\bibnamefont {Cuniberti}},\ }\bibfield  {title} {\enquote {\bibinfo {title} {Spin-selective transport through helical molecular systems},}\ }\href {\doibase 10.1103/PhysRevB.85.081404} {\bibfield  {journal} {\bibinfo  {journal} {Phys. Rev. B}\ }\textbf {\bibinfo {volume} {85}},\ \bibinfo {pages} {081404} (\bibinfo {year} {2012})}\BibitemShut {NoStop}%
\bibitem [{\citenamefont {Guo}\ and\ \citenamefont {Sun}(2012)}]{Guo2012}%
  \BibitemOpen
  \bibfield  {author} {\bibinfo {author} {\bibfnamefont {A.-M.}\ \bibnamefont {Guo}}\ and\ \bibinfo {author} {\bibfnamefont {Q.-f.}\ \bibnamefont {Sun}},\ }\bibfield  {title} {\enquote {\bibinfo {title} {Spin-selective transport of electrons in dna double helix},}\ }\href {\doibase 10.1103/PhysRevLett.108.218102} {\bibfield  {journal} {\bibinfo  {journal} {Phys. Rev. Lett.}\ }\textbf {\bibinfo {volume} {108}},\ \bibinfo {pages} {218102} (\bibinfo {year} {2012})}\BibitemShut {NoStop}%
\bibitem [{\citenamefont {Medina}\ \emph {et~al.}(2012)\citenamefont {Medina}, \citenamefont {L{\'{o}}pez}, \citenamefont {Ratner},\ and\ \citenamefont {Mujica}}]{Medina2012}%
  \BibitemOpen
  \bibfield  {author} {\bibinfo {author} {\bibfnamefont {E.}~\bibnamefont {Medina}}, \bibinfo {author} {\bibfnamefont {F.}~\bibnamefont {L{\'{o}}pez}}, \bibinfo {author} {\bibfnamefont {M.~A.}\ \bibnamefont {Ratner}}, \ and\ \bibinfo {author} {\bibfnamefont {V.}~\bibnamefont {Mujica}},\ }\bibfield  {title} {\enquote {\bibinfo {title} {Chiral molecular films as electron polarizers and polarization modulators},}\ }\href {\doibase 10.1209/0295-5075/99/17006} {\bibfield  {journal} {\bibinfo  {journal} {{EPL} (Europhysics Letters)}\ }\textbf {\bibinfo {volume} {99}},\ \bibinfo {pages} {17006} (\bibinfo {year} {2012})}\BibitemShut {NoStop}%
\bibitem [{\citenamefont {Gersten}, \citenamefont {Kaasbjerg},\ and\ \citenamefont {Nitzan}(2013)}]{Gersten2013}%
  \BibitemOpen
  \bibfield  {author} {\bibinfo {author} {\bibfnamefont {J.}~\bibnamefont {Gersten}}, \bibinfo {author} {\bibfnamefont {K.}~\bibnamefont {Kaasbjerg}}, \ and\ \bibinfo {author} {\bibfnamefont {A.}~\bibnamefont {Nitzan}},\ }\bibfield  {title} {\enquote {\bibinfo {title} {Induced spin filtering in electron transmission through chiral molecular layers adsorbed on metals with strong spin-orbit coupling},}\ }\href {\doibase 10.1063/1.4820907} {\bibfield  {journal} {\bibinfo  {journal} {The Journal of Chemical Physics}\ }\textbf {\bibinfo {volume} {139}},\ \bibinfo {pages} {114111} (\bibinfo {year} {2013})}\BibitemShut {NoStop}%
\bibitem [{\citenamefont {Guo}\ and\ \citenamefont {Sun}(2014)}]{Guo2014}%
  \BibitemOpen
  \bibfield  {author} {\bibinfo {author} {\bibfnamefont {A.-M.}\ \bibnamefont {Guo}}\ and\ \bibinfo {author} {\bibfnamefont {Q.-F.}\ \bibnamefont {Sun}},\ }\bibfield  {title} {\enquote {\bibinfo {title} {Spin-dependent electron transport in protein-like single-helical molecules},}\ }\href {\doibase 10.1073/pnas.1407716111} {\bibfield  {journal} {\bibinfo  {journal} {Proceedings of the National Academy of Sciences}\ }\textbf {\bibinfo {volume} {111}},\ \bibinfo {pages} {11658--11662} (\bibinfo {year} {2014})}\BibitemShut {NoStop}%
\bibitem [{\citenamefont {Matityahu}\ \emph {et~al.}(2016)\citenamefont {Matityahu}, \citenamefont {Utsumi}, \citenamefont {Aharony}, \citenamefont {Entin-Wohlman},\ and\ \citenamefont {Balseiro}}]{Matityahu2016}%
  \BibitemOpen
  \bibfield  {author} {\bibinfo {author} {\bibfnamefont {S.}~\bibnamefont {Matityahu}}, \bibinfo {author} {\bibfnamefont {Y.}~\bibnamefont {Utsumi}}, \bibinfo {author} {\bibfnamefont {A.}~\bibnamefont {Aharony}}, \bibinfo {author} {\bibfnamefont {O.}~\bibnamefont {Entin-Wohlman}}, \ and\ \bibinfo {author} {\bibfnamefont {C.~A.}\ \bibnamefont {Balseiro}},\ }\bibfield  {title} {\enquote {\bibinfo {title} {Spin-dependent transport through a chiral molecule in the presence of spin-orbit interaction and nonunitary effects},}\ }\href {\doibase 10.1103/PhysRevB.93.075407} {\bibfield  {journal} {\bibinfo  {journal} {Phys. Rev. B}\ }\textbf {\bibinfo {volume} {93}},\ \bibinfo {pages} {075407} (\bibinfo {year} {2016})}\BibitemShut {NoStop}%
\bibitem [{\citenamefont {Diaz}\ \emph {et~al.}(2018)\citenamefont {Diaz}, \citenamefont {Albares}, \citenamefont {Estevez}, \citenamefont {Cerveró}, \citenamefont {Gaul}, \citenamefont {Diez},\ and\ \citenamefont {Domínguez-Adame}}]{Diaz_2018}%
  \BibitemOpen
  \bibfield  {author} {\bibinfo {author} {\bibfnamefont {E.}~\bibnamefont {Diaz}}, \bibinfo {author} {\bibfnamefont {P.}~\bibnamefont {Albares}}, \bibinfo {author} {\bibfnamefont {P.~G.}\ \bibnamefont {Estevez}}, \bibinfo {author} {\bibfnamefont {J.~M.}\ \bibnamefont {Cerveró}}, \bibinfo {author} {\bibfnamefont {C.}~\bibnamefont {Gaul}}, \bibinfo {author} {\bibfnamefont {E.}~\bibnamefont {Diez}}, \ and\ \bibinfo {author} {\bibfnamefont {F.}~\bibnamefont {Domínguez-Adame}},\ }\bibfield  {title} {\enquote {\bibinfo {title} {Spin dynamics in helical molecules with nonlinear interactions},}\ }\href {\doibase 10.1088/1367-2630/aabb91} {\bibfield  {journal} {\bibinfo  {journal} {New Journal of Physics}\ }\textbf {\bibinfo {volume} {20}},\ \bibinfo {pages} {043055} (\bibinfo {year} {2018})}\BibitemShut {NoStop}%
\bibitem [{\citenamefont {Fransson}(2019)}]{Fransson2019}%
  \BibitemOpen
  \bibfield  {author} {\bibinfo {author} {\bibfnamefont {J.}~\bibnamefont {Fransson}},\ }\bibfield  {title} {\enquote {\bibinfo {title} {Chirality-induced spin selectivity: The role of electron correlations},}\ }\href {\doibase 10.1021/acs.jpclett.9b02929} {\bibfield  {journal} {\bibinfo  {journal} {The Journal of Physical Chemistry Letters}\ }\textbf {\bibinfo {volume} {10}},\ \bibinfo {pages} {7126--7132} (\bibinfo {year} {2019})}\BibitemShut {NoStop}%
\bibitem [{\citenamefont {Michaeli}\ and\ \citenamefont {Naaman}(2019)}]{Michaeli2019}%
  \BibitemOpen
  \bibfield  {author} {\bibinfo {author} {\bibfnamefont {K.}~\bibnamefont {Michaeli}}\ and\ \bibinfo {author} {\bibfnamefont {R.}~\bibnamefont {Naaman}},\ }\bibfield  {title} {\enquote {\bibinfo {title} {Origin of spin-dependent tunneling through chiral molecules},}\ }\href {\doibase 10.1021/acs.jpcc.9b05020} {\bibfield  {journal} {\bibinfo  {journal} {The Journal of Physical Chemistry C}\ }\textbf {\bibinfo {volume} {123}},\ \bibinfo {pages} {17043--17048} (\bibinfo {year} {2019})}\BibitemShut {NoStop}%
\bibitem [{\citenamefont {Dalum}\ and\ \citenamefont {Hedegård}(2019)}]{Dalum2019}%
  \BibitemOpen
  \bibfield  {author} {\bibinfo {author} {\bibfnamefont {S.}~\bibnamefont {Dalum}}\ and\ \bibinfo {author} {\bibfnamefont {P.}~\bibnamefont {Hedegård}},\ }\bibfield  {title} {\enquote {\bibinfo {title} {Theory of chiral induced spin selectivity},}\ }\href {\doibase 10.1021/acs.nanolett.9b01707} {\bibfield  {journal} {\bibinfo  {journal} {Nano Letters}\ }\textbf {\bibinfo {volume} {19}},\ \bibinfo {pages} {5253--5259} (\bibinfo {year} {2019})},\ \bibinfo {note} {pMID: 31265313}\BibitemShut {NoStop}%
\bibitem [{\citenamefont {Fransson}(2020)}]{Fransson2020}%
  \BibitemOpen
  \bibfield  {author} {\bibinfo {author} {\bibfnamefont {J.}~\bibnamefont {Fransson}},\ }\bibfield  {title} {\enquote {\bibinfo {title} {Vibrational origin of exchange splitting and ''chiral-induced spin selectivity},}\ }\href {\doibase 10.1103/PhysRevB.102.235416} {\bibfield  {journal} {\bibinfo  {journal} {Phys. Rev. B}\ }\textbf {\bibinfo {volume} {102}},\ \bibinfo {pages} {235416} (\bibinfo {year} {2020})}\BibitemShut {NoStop}%
\bibitem [{\citenamefont {Ghazaryan}, \citenamefont {Lemeshko},\ and\ \citenamefont {Volosniev}(2020)}]{Ghazaryan2020}%
  \BibitemOpen
  \bibfield  {author} {\bibinfo {author} {\bibfnamefont {A.}~\bibnamefont {Ghazaryan}}, \bibinfo {author} {\bibfnamefont {M.}~\bibnamefont {Lemeshko}}, \ and\ \bibinfo {author} {\bibfnamefont {A.~G.}\ \bibnamefont {Volosniev}},\ }\bibfield  {title} {\enquote {\bibinfo {title} {Filtering spins by scattering from a lattice of point magnets},}\ }\href {\doibase 10.1038/s42005-020-00445-8} {\bibfield  {journal} {\bibinfo  {journal} {Communications Physics}\ }\textbf {\bibinfo {volume} {3}} (\bibinfo {year} {2020}),\ 10.1038/s42005-020-00445-8}\BibitemShut {NoStop}%
\bibitem [{\citenamefont {Ghazaryan}, \citenamefont {Paltiel},\ and\ \citenamefont {Lemeshko}(2020)}]{Ghazaryan2020a}%
  \BibitemOpen
  \bibfield  {author} {\bibinfo {author} {\bibfnamefont {A.}~\bibnamefont {Ghazaryan}}, \bibinfo {author} {\bibfnamefont {Y.}~\bibnamefont {Paltiel}}, \ and\ \bibinfo {author} {\bibfnamefont {M.}~\bibnamefont {Lemeshko}},\ }\bibfield  {title} {\enquote {\bibinfo {title} {Analytic model of chiral-induced spin selectivity},}\ }\href {\doibase 10.1021/acs.jpcc.0c02584} {\bibfield  {journal} {\bibinfo  {journal} {The Journal of Physical Chemistry C}\ }\textbf {\bibinfo {volume} {124}},\ \bibinfo {pages} {11716--11721} (\bibinfo {year} {2020})},\ \bibinfo {note} {pMID: 32499842}\BibitemShut {NoStop}%
\bibitem [{\citenamefont {Alwan}\ and\ \citenamefont {Dubi}(2021)}]{Alwan2021}%
  \BibitemOpen
  \bibfield  {author} {\bibinfo {author} {\bibfnamefont {S.}~\bibnamefont {Alwan}}\ and\ \bibinfo {author} {\bibfnamefont {Y.}~\bibnamefont {Dubi}},\ }\bibfield  {title} {\enquote {\bibinfo {title} {Spinterface origin for the chirality-induced spin-selectivity effect},}\ }\href {\doibase 10.1021/jacs.1c05637} {\bibfield  {journal} {\bibinfo  {journal} {Journal of the American Chemical Society}\ }\textbf {\bibinfo {volume} {143}},\ \bibinfo {pages} {14235--14241} (\bibinfo {year} {2021})},\ \bibinfo {note} {pMID: 34460242}\BibitemShut {NoStop}%
\bibitem [{\citenamefont {Liu}\ \emph {et~al.}(2021)\citenamefont {Liu}, \citenamefont {Xiao}, \citenamefont {Koo},\ and\ \citenamefont {Yan}}]{Liu2021}%
  \BibitemOpen
  \bibfield  {author} {\bibinfo {author} {\bibfnamefont {Y.}~\bibnamefont {Liu}}, \bibinfo {author} {\bibfnamefont {J.}~\bibnamefont {Xiao}}, \bibinfo {author} {\bibfnamefont {J.}~\bibnamefont {Koo}}, \ and\ \bibinfo {author} {\bibfnamefont {B.}~\bibnamefont {Yan}},\ }\bibfield  {title} {\enquote {\bibinfo {title} {Chirality-driven topological electronic structure of {DNA}-like materials},}\ }\href {\doibase 10.1038/s41563-021-00924-5} {\bibfield  {journal} {\bibinfo  {journal} {Nature Materials}\ }\textbf {\bibinfo {volume} {20}},\ \bibinfo {pages} {638--644} (\bibinfo {year} {2021})}\BibitemShut {NoStop}%
\bibitem [{\citenamefont {Wolf}\ \emph {et~al.}(2022)\citenamefont {Wolf}, \citenamefont {Liu}, \citenamefont {Xiao}, \citenamefont {Park},\ and\ \citenamefont {Yan}}]{Wolf2022}%
  \BibitemOpen
  \bibfield  {author} {\bibinfo {author} {\bibfnamefont {Y.}~\bibnamefont {Wolf}}, \bibinfo {author} {\bibfnamefont {Y.}~\bibnamefont {Liu}}, \bibinfo {author} {\bibfnamefont {J.}~\bibnamefont {Xiao}}, \bibinfo {author} {\bibfnamefont {N.}~\bibnamefont {Park}}, \ and\ \bibinfo {author} {\bibfnamefont {B.}~\bibnamefont {Yan}},\ }\bibfield  {title} {\enquote {\bibinfo {title} {Unusual spin polarization in the chirality-induced spin selectivity},}\ }\href {\doibase 10.1021/acsnano.2c07088} {\bibfield  {journal} {\bibinfo  {journal} {ACS Nano}\ }\textbf {\bibinfo {volume} {16}},\ \bibinfo {pages} {18601--18607} (\bibinfo {year} {2022})},\ \bibinfo {note} {pMID: 36282509}\BibitemShut {NoStop}%
\bibitem [{\citenamefont {Vittmann}\ \emph {et~al.}(2023)\citenamefont {Vittmann}, \citenamefont {Lim}, \citenamefont {Tamascelli}, \citenamefont {Huelga},\ and\ \citenamefont {Plenio}}]{Vittmann2023}%
  \BibitemOpen
  \bibfield  {author} {\bibinfo {author} {\bibfnamefont {C.}~\bibnamefont {Vittmann}}, \bibinfo {author} {\bibfnamefont {J.}~\bibnamefont {Lim}}, \bibinfo {author} {\bibfnamefont {D.}~\bibnamefont {Tamascelli}}, \bibinfo {author} {\bibfnamefont {S.~F.}\ \bibnamefont {Huelga}}, \ and\ \bibinfo {author} {\bibfnamefont {M.~B.}\ \bibnamefont {Plenio}},\ }\bibfield  {title} {\enquote {\bibinfo {title} {Spin-dependent momentum conservation of electron–phonon scattering in chirality-induced spin selectivity},}\ }\href {\doibase 10.1021/acs.jpclett.2c03224} {\bibfield  {journal} {\bibinfo  {journal} {The Journal of Physical Chemistry Letters}\ }\textbf {\bibinfo {volume} {14}},\ \bibinfo {pages} {340--346} (\bibinfo {year} {2023})},\ \bibinfo {note} {pMID: 36625481}\BibitemShut {NoStop}%
\bibitem [{\citenamefont {Sanvito}(2010)}]{Sanvito2010}%
  \BibitemOpen
  \bibfield  {author} {\bibinfo {author} {\bibfnamefont {S.}~\bibnamefont {Sanvito}},\ }\bibfield  {title} {\enquote {\bibinfo {title} {The rise of spinterface science},}\ }\href {\doibase 10.1038/nphys1714} {\bibfield  {journal} {\bibinfo  {journal} {Nature Physics}\ }\textbf {\bibinfo {volume} {6}},\ \bibinfo {pages} {562--564} (\bibinfo {year} {2010})}\BibitemShut {NoStop}%
\bibitem [{\citenamefont {Geyer}\ \emph {et~al.}(2019)\citenamefont {Geyer}, \citenamefont {Gutierrez}, \citenamefont {Mujica},\ and\ \citenamefont {Cuniberti}}]{Geyer2019}%
  \BibitemOpen
  \bibfield  {author} {\bibinfo {author} {\bibfnamefont {M.}~\bibnamefont {Geyer}}, \bibinfo {author} {\bibfnamefont {R.}~\bibnamefont {Gutierrez}}, \bibinfo {author} {\bibfnamefont {V.}~\bibnamefont {Mujica}}, \ and\ \bibinfo {author} {\bibfnamefont {G.}~\bibnamefont {Cuniberti}},\ }\bibfield  {title} {\enquote {\bibinfo {title} {Chirality-induced spin selectivity in a coarse-grained tight-binding model for helicene},}\ }\href {\doibase 10.1021/acs.jpcc.9b07764} {\bibfield  {journal} {\bibinfo  {journal} {The Journal of Physical Chemistry C}\ }\textbf {\bibinfo {volume} {123}},\ \bibinfo {pages} {27230--27241} (\bibinfo {year} {2019})}\BibitemShut {NoStop}%
\bibitem [{\citenamefont {Volosniev}\ \emph {et~al.}(2021)\citenamefont {Volosniev}, \citenamefont {Alpern}, \citenamefont {Paltiel}, \citenamefont {Millo}, \citenamefont {Lemeshko},\ and\ \citenamefont {Ghazaryan}}]{Volosniev2021}%
  \BibitemOpen
  \bibfield  {author} {\bibinfo {author} {\bibfnamefont {A.~G.}\ \bibnamefont {Volosniev}}, \bibinfo {author} {\bibfnamefont {H.}~\bibnamefont {Alpern}}, \bibinfo {author} {\bibfnamefont {Y.}~\bibnamefont {Paltiel}}, \bibinfo {author} {\bibfnamefont {O.}~\bibnamefont {Millo}}, \bibinfo {author} {\bibfnamefont {M.}~\bibnamefont {Lemeshko}}, \ and\ \bibinfo {author} {\bibfnamefont {A.}~\bibnamefont {Ghazaryan}},\ }\bibfield  {title} {\enquote {\bibinfo {title} {Interplay between friction and spin-orbit coupling as a source of spin polarization},}\ }\href {\doibase 10.1103/PhysRevB.104.024430} {\bibfield  {journal} {\bibinfo  {journal} {Phys. Rev. B}\ }\textbf {\bibinfo {volume} {104}},\ \bibinfo {pages} {024430} (\bibinfo {year} {2021})}\BibitemShut {NoStop}%
\bibitem [{\citenamefont {Eerenstein}, \citenamefont {Mathur},\ and\ \citenamefont {Scott}(2006)}]{Eerenstein2006}%
  \BibitemOpen
  \bibfield  {author} {\bibinfo {author} {\bibfnamefont {W.}~\bibnamefont {Eerenstein}}, \bibinfo {author} {\bibfnamefont {N.~D.}\ \bibnamefont {Mathur}}, \ and\ \bibinfo {author} {\bibfnamefont {J.~F.}\ \bibnamefont {Scott}},\ }\bibfield  {title} {\enquote {\bibinfo {title} {Multiferroic and magnetoelectric materials},}\ }\href {\doibase 10.1038/nature05023} {\bibfield  {journal} {\bibinfo  {journal} {Nature}\ }\textbf {\bibinfo {volume} {442}},\ \bibinfo {pages} {759--765} (\bibinfo {year} {2006})}\BibitemShut {NoStop}%
\bibitem [{\citenamefont {Niranjan}\ \emph {et~al.}(2010)\citenamefont {Niranjan}, \citenamefont {Duan}, \citenamefont {Jaswal},\ and\ \citenamefont {Tsymbal}}]{Niranjan2010}%
  \BibitemOpen
  \bibfield  {author} {\bibinfo {author} {\bibfnamefont {M.~K.}\ \bibnamefont {Niranjan}}, \bibinfo {author} {\bibfnamefont {C.-G.}\ \bibnamefont {Duan}}, \bibinfo {author} {\bibfnamefont {S.~S.}\ \bibnamefont {Jaswal}}, \ and\ \bibinfo {author} {\bibfnamefont {E.~Y.}\ \bibnamefont {Tsymbal}},\ }\bibfield  {title} {\enquote {\bibinfo {title} {Electric field effect on magnetization at the fe/{MgO}(001) interface},}\ }\href {\doibase 10.1063/1.3443658} {\bibfield  {journal} {\bibinfo  {journal} {Applied Physics Letters}\ }\textbf {\bibinfo {volume} {96}},\ \bibinfo {pages} {222504} (\bibinfo {year} {2010})}\BibitemShut {NoStop}%
\bibitem [{\citenamefont {Kumari}\ and\ \citenamefont {Niranjan}(2021)}]{KUMARI2021}%
  \BibitemOpen
  \bibfield  {author} {\bibinfo {author} {\bibfnamefont {P.~K.}\ \bibnamefont {Kumari}}\ and\ \bibinfo {author} {\bibfnamefont {M.~K.}\ \bibnamefont {Niranjan}},\ }\bibfield  {title} {\enquote {\bibinfo {title} {Interface magnetoelectric effect and its sensitivity on interface structures in fe/agnbo3 and srruo3/agnbo3 heterostructures: A first-principles investigation},}\ }\href {\doibase https://doi.org/10.1016/j.jmmm.2020.167372} {\bibfield  {journal} {\bibinfo  {journal} {Journal of Magnetism and Magnetic Materials}\ }\textbf {\bibinfo {volume} {517}},\ \bibinfo {pages} {167372} (\bibinfo {year} {2021})}\BibitemShut {NoStop}%
\bibitem [{\citenamefont {Cinchetti}, \citenamefont {Dediu},\ and\ \citenamefont {Hueso}(2017)}]{Cinchetti2017}%
  \BibitemOpen
  \bibfield  {author} {\bibinfo {author} {\bibfnamefont {M.}~\bibnamefont {Cinchetti}}, \bibinfo {author} {\bibfnamefont {V.~A.}\ \bibnamefont {Dediu}}, \ and\ \bibinfo {author} {\bibfnamefont {L.~E.}\ \bibnamefont {Hueso}},\ }\bibfield  {title} {\enquote {\bibinfo {title} {Activating the molecular spinterface},}\ }\href {\doibase 10.1038/nmat4902} {\bibfield  {journal} {\bibinfo  {journal} {Nature Materials}\ }\textbf {\bibinfo {volume} {16}},\ \bibinfo {pages} {507--515} (\bibinfo {year} {2017})}\BibitemShut {NoStop}%
\bibitem [{\citenamefont {Soumyanarayanan}\ \emph {et~al.}(2016)\citenamefont {Soumyanarayanan}, \citenamefont {Reyren}, \citenamefont {Fert},\ and\ \citenamefont {Panagopoulos}}]{Soumyanarayanan2016}%
  \BibitemOpen
  \bibfield  {author} {\bibinfo {author} {\bibfnamefont {A.}~\bibnamefont {Soumyanarayanan}}, \bibinfo {author} {\bibfnamefont {N.}~\bibnamefont {Reyren}}, \bibinfo {author} {\bibfnamefont {A.}~\bibnamefont {Fert}}, \ and\ \bibinfo {author} {\bibfnamefont {C.}~\bibnamefont {Panagopoulos}},\ }\bibfield  {title} {\enquote {\bibinfo {title} {Emergent phenomena induced by spin{\textendash}orbit coupling at surfaces and interfaces},}\ }\href {\doibase 10.1038/nature19820} {\bibfield  {journal} {\bibinfo  {journal} {Nature}\ }\textbf {\bibinfo {volume} {539}},\ \bibinfo {pages} {509--517} (\bibinfo {year} {2016})}\BibitemShut {NoStop}%
\bibitem [{\citenamefont {Callsen}\ \emph {et~al.}(2013)\citenamefont {Callsen}, \citenamefont {Caciuc}, \citenamefont {Kiselev}, \citenamefont {Atodiresei},\ and\ \citenamefont {Bl\"ugel}}]{callsen2013}%
  \BibitemOpen
  \bibfield  {author} {\bibinfo {author} {\bibfnamefont {M.}~\bibnamefont {Callsen}}, \bibinfo {author} {\bibfnamefont {V.}~\bibnamefont {Caciuc}}, \bibinfo {author} {\bibfnamefont {N.}~\bibnamefont {Kiselev}}, \bibinfo {author} {\bibfnamefont {N.}~\bibnamefont {Atodiresei}}, \ and\ \bibinfo {author} {\bibfnamefont {S.}~\bibnamefont {Bl\"ugel}},\ }\bibfield  {title} {\enquote {\bibinfo {title} {Magnetic hardening induced by nonmagnetic organic molecules},}\ }\href {\doibase 10.1103/PhysRevLett.111.106805} {\bibfield  {journal} {\bibinfo  {journal} {Phys. Rev. Lett.}\ }\textbf {\bibinfo {volume} {111}},\ \bibinfo {pages} {106805} (\bibinfo {year} {2013})}\BibitemShut {NoStop}%
\bibitem [{\citenamefont {Bairagi}\ \emph {et~al.}(2015)\citenamefont {Bairagi}, \citenamefont {Bellec}, \citenamefont {Repain}, \citenamefont {Chacon}, \citenamefont {Girard}, \citenamefont {Garreau}, \citenamefont {Lagoute}, \citenamefont {Rousset}, \citenamefont {Breitwieser}, \citenamefont {Hu}, \citenamefont {Chao}, \citenamefont {Pai}, \citenamefont {Li}, \citenamefont {Smogunov},\ and\ \citenamefont {Barreteau}}]{Bairagi2015}%
  \BibitemOpen
  \bibfield  {author} {\bibinfo {author} {\bibfnamefont {K.}~\bibnamefont {Bairagi}}, \bibinfo {author} {\bibfnamefont {A.}~\bibnamefont {Bellec}}, \bibinfo {author} {\bibfnamefont {V.}~\bibnamefont {Repain}}, \bibinfo {author} {\bibfnamefont {C.}~\bibnamefont {Chacon}}, \bibinfo {author} {\bibfnamefont {Y.}~\bibnamefont {Girard}}, \bibinfo {author} {\bibfnamefont {Y.}~\bibnamefont {Garreau}}, \bibinfo {author} {\bibfnamefont {J.}~\bibnamefont {Lagoute}}, \bibinfo {author} {\bibfnamefont {S.}~\bibnamefont {Rousset}}, \bibinfo {author} {\bibfnamefont {R.}~\bibnamefont {Breitwieser}}, \bibinfo {author} {\bibfnamefont {Y.-C.}\ \bibnamefont {Hu}}, \bibinfo {author} {\bibfnamefont {Y.~C.}\ \bibnamefont {Chao}}, \bibinfo {author} {\bibfnamefont {W.~W.}\ \bibnamefont {Pai}}, \bibinfo {author} {\bibfnamefont {D.}~\bibnamefont {Li}}, \bibinfo {author} {\bibfnamefont {A.}~\bibnamefont {Smogunov}}, \ and\ \bibinfo {author} {\bibfnamefont {C.}~\bibnamefont {Barreteau}},\ }\bibfield  {title} {\enquote {\bibinfo {title} {Tuning the magnetic anisotropy at a molecule-metal interface},}\ }\href {\doibase 10.1103/PhysRevLett.114.247203} {\bibfield  {journal} {\bibinfo  {journal} {Phys. Rev. Lett.}\ }\textbf {\bibinfo {volume} {114}},\ \bibinfo {pages} {247203} (\bibinfo {year} {2015})}\BibitemShut {NoStop}%
\bibitem [{\citenamefont {Zhang}(1999)}]{Zhang1999}%
  \BibitemOpen
  \bibfield  {author} {\bibinfo {author} {\bibfnamefont {S.}~\bibnamefont {Zhang}},\ }\bibfield  {title} {\enquote {\bibinfo {title} {Spin-dependent surface screening in ferromagnets and magnetic tunnel junctions},}\ }\href {\doibase 10.1103/PhysRevLett.83.640} {\bibfield  {journal} {\bibinfo  {journal} {Phys. Rev. Lett.}\ }\textbf {\bibinfo {volume} {83}},\ \bibinfo {pages} {640--643} (\bibinfo {year} {1999})}\BibitemShut {NoStop}%
\bibitem [{\citenamefont {Duan}\ \emph {et~al.}(2008)\citenamefont {Duan}, \citenamefont {Velev}, \citenamefont {Sabirianov}, \citenamefont {Zhu}, \citenamefont {Chu}, \citenamefont {Jaswal},\ and\ \citenamefont {Tsymbal}}]{Duan2008}%
  \BibitemOpen
  \bibfield  {author} {\bibinfo {author} {\bibfnamefont {C.-G.}\ \bibnamefont {Duan}}, \bibinfo {author} {\bibfnamefont {J.~P.}\ \bibnamefont {Velev}}, \bibinfo {author} {\bibfnamefont {R.~F.}\ \bibnamefont {Sabirianov}}, \bibinfo {author} {\bibfnamefont {Z.}~\bibnamefont {Zhu}}, \bibinfo {author} {\bibfnamefont {J.}~\bibnamefont {Chu}}, \bibinfo {author} {\bibfnamefont {S.~S.}\ \bibnamefont {Jaswal}}, \ and\ \bibinfo {author} {\bibfnamefont {E.~Y.}\ \bibnamefont {Tsymbal}},\ }\bibfield  {title} {\enquote {\bibinfo {title} {Surface magnetoelectric effect in ferromagnetic metal films},}\ }\href {\doibase 10.1103/PhysRevLett.101.137201} {\bibfield  {journal} {\bibinfo  {journal} {Phys. Rev. Lett.}\ }\textbf {\bibinfo {volume} {101}},\ \bibinfo {pages} {137201} (\bibinfo {year} {2008})}\BibitemShut {NoStop}%
\bibitem [{\citenamefont {Natan}\ \emph {et~al.}(2007)\citenamefont {Natan}, \citenamefont {Kronik}, \citenamefont {Haick},\ and\ \citenamefont {Tung}}]{Natan2007}%
  \BibitemOpen
  \bibfield  {author} {\bibinfo {author} {\bibfnamefont {A.}~\bibnamefont {Natan}}, \bibinfo {author} {\bibfnamefont {L.}~\bibnamefont {Kronik}}, \bibinfo {author} {\bibfnamefont {H.}~\bibnamefont {Haick}}, \ and\ \bibinfo {author} {\bibfnamefont {R.}~\bibnamefont {Tung}},\ }\bibfield  {title} {\enquote {\bibinfo {title} {Electrostatic properties of ideal and non-ideal polar organic monolayers: Implications for electronic devices},}\ }\href {\doibase 10.1002/adma.200701681} {\bibfield  {journal} {\bibinfo  {journal} {Advanced Materials}\ }\textbf {\bibinfo {volume} {19}},\ \bibinfo {pages} {4103--4117} (\bibinfo {year} {2007})}\BibitemShut {NoStop}%
\bibitem [{\citenamefont {Monti}(2012)}]{Monti2012}%
  \BibitemOpen
  \bibfield  {author} {\bibinfo {author} {\bibfnamefont {O.~L.~A.}\ \bibnamefont {Monti}},\ }\bibfield  {title} {\enquote {\bibinfo {title} {Understanding interfacial electronic structure and charge transfer: An electrostatic perspective},}\ }\href {\doibase 10.1021/jz300850x} {\bibfield  {journal} {\bibinfo  {journal} {The Journal of Physical Chemistry Letters}\ }\textbf {\bibinfo {volume} {3}},\ \bibinfo {pages} {2342--2351} (\bibinfo {year} {2012})}\BibitemShut {NoStop}%
\bibitem [{\citenamefont {Gold}(1974)}]{Gold1974}%
  \BibitemOpen
  \bibfield  {author} {\bibinfo {author} {\bibfnamefont {A.~V.}\ \bibnamefont {Gold}},\ }\bibfield  {title} {\enquote {\bibinfo {title} {Review paper: Fermi surfaces of the ferromagnetic transition metals},}\ }\href {\doibase 10.1007/bf00655857} {\bibfield  {journal} {\bibinfo  {journal} {Journal of Low Temperature Physics}\ }\textbf {\bibinfo {volume} {16}},\ \bibinfo {pages} {3--42} (\bibinfo {year} {1974})}\BibitemShut {NoStop}%
\bibitem [{\citenamefont {Rashba}(1960)}]{Rashba1960}%
  \BibitemOpen
  \bibfield  {author} {\bibinfo {author} {\bibfnamefont {E.}~\bibnamefont {Rashba}},\ }\href@noop {} {\bibfield  {journal} {\bibinfo  {journal} {Sov. Phys.-Solid State}\ }\textbf {\bibinfo {volume} {2}},\ \bibinfo {pages} {1109} (\bibinfo {year} {1960})}\BibitemShut {NoStop}%
\bibitem [{\citenamefont {{Bychkov}}\ and\ \citenamefont {{Rashba}}(1984)}]{rashba-1984}%
  \BibitemOpen
  \bibfield  {author} {\bibinfo {author} {\bibfnamefont {Y.~A.}\ \bibnamefont {{Bychkov}}}\ and\ \bibinfo {author} {\bibfnamefont {{\'E}.~I.}\ \bibnamefont {{Rashba}}},\ }\bibfield  {title} {\enquote {\bibinfo {title} {{Properties of a 2D electron gas with lifted spectral degeneracy}},}\ }\href@noop {} {\bibfield  {journal} {\bibinfo  {journal} {Soviet Journal of Experimental and Theoretical Physics Letters}\ }\textbf {\bibinfo {volume} {39}},\ \bibinfo {pages} {78} (\bibinfo {year} {1984})}\BibitemShut {NoStop}%
\bibitem [{\citenamefont {Manchon}\ \emph {et~al.}(2015)\citenamefont {Manchon}, \citenamefont {Koo}, \citenamefont {Nitta}, \citenamefont {Frolov},\ and\ \citenamefont {Duine}}]{manchon-2015}%
  \BibitemOpen
  \bibfield  {author} {\bibinfo {author} {\bibfnamefont {A.}~\bibnamefont {Manchon}}, \bibinfo {author} {\bibfnamefont {H.~C.}\ \bibnamefont {Koo}}, \bibinfo {author} {\bibfnamefont {J.}~\bibnamefont {Nitta}}, \bibinfo {author} {\bibfnamefont {S.~M.}\ \bibnamefont {Frolov}}, \ and\ \bibinfo {author} {\bibfnamefont {R.~A.}\ \bibnamefont {Duine}},\ }\bibfield  {title} {\enquote {\bibinfo {title} {New perspectives for rashba spin{\textendash}orbit coupling},}\ }\href {\doibase 10.1038/nmat4360} {\bibfield  {journal} {\bibinfo  {journal} {Nature Materials}\ }\textbf {\bibinfo {volume} {14}},\ \bibinfo {pages} {871--882} (\bibinfo {year} {2015})}\BibitemShut {NoStop}%
\bibitem [{\citenamefont {Meijer}, \citenamefont {Morpurgo},\ and\ \citenamefont {Klapwijk}(2002)}]{meijer-2002}%
  \BibitemOpen
  \bibfield  {author} {\bibinfo {author} {\bibfnamefont {F.~E.}\ \bibnamefont {Meijer}}, \bibinfo {author} {\bibfnamefont {A.~F.}\ \bibnamefont {Morpurgo}}, \ and\ \bibinfo {author} {\bibfnamefont {T.~M.}\ \bibnamefont {Klapwijk}},\ }\bibfield  {title} {\enquote {\bibinfo {title} {One-dimensional ring in the presence of rashba spin-orbit interaction: Derivation of the correct hamiltonian},}\ }\href {\doibase 10.1103/PhysRevB.66.033107} {\bibfield  {journal} {\bibinfo  {journal} {Phys. Rev. B}\ }\textbf {\bibinfo {volume} {66}},\ \bibinfo {pages} {033107} (\bibinfo {year} {2002})}\BibitemShut {NoStop}%
\bibitem [{\citenamefont {Berche}, \citenamefont {Chatelain},\ and\ \citenamefont {Medina}(2010)}]{berche-2010}%
  \BibitemOpen
  \bibfield  {author} {\bibinfo {author} {\bibfnamefont {B.}~\bibnamefont {Berche}}, \bibinfo {author} {\bibfnamefont {C.}~\bibnamefont {Chatelain}}, \ and\ \bibinfo {author} {\bibfnamefont {E.}~\bibnamefont {Medina}},\ }\bibfield  {title} {\enquote {\bibinfo {title} {Mesoscopic rings with spin-orbit interactions},}\ }\href {\doibase 10.1088/0143-0807/31/5/026} {\bibfield  {journal} {\bibinfo  {journal} {European Journal of Physics}\ }\textbf {\bibinfo {volume} {31}},\ \bibinfo {pages} {1267} (\bibinfo {year} {2010})}\BibitemShut {NoStop}%
\bibitem [{\citenamefont {Stoner}\ and\ \citenamefont {Wohlfarth}(1948)}]{Stoner1948}%
  \BibitemOpen
  \bibfield  {author} {\bibinfo {author} {\bibfnamefont {E.~C.}\ \bibnamefont {Stoner}}\ and\ \bibinfo {author} {\bibfnamefont {E.~P.}\ \bibnamefont {Wohlfarth}},\ }\bibfield  {title} {\enquote {\bibinfo {title} {A mechanism of magnetic hysteresis in heterogeneous alloys},}\ }\href {\doibase 10.1098/rsta.1948.0007} {\bibfield  {journal} {\bibinfo  {journal} {Philosophical Transactions of the Royal Society of London. Series A, Mathematical and Physical Sciences}\ }\textbf {\bibinfo {volume} {240}},\ \bibinfo {pages} {599--642} (\bibinfo {year} {1948})}\BibitemShut {NoStop}%
\bibitem [{\citenamefont {Tannous}\ and\ \citenamefont {Gieraltowski}(2008)}]{Tannous_2008}%
  \BibitemOpen
  \bibfield  {author} {\bibinfo {author} {\bibfnamefont {C.}~\bibnamefont {Tannous}}\ and\ \bibinfo {author} {\bibfnamefont {J.}~\bibnamefont {Gieraltowski}},\ }\bibfield  {title} {\enquote {\bibinfo {title} {The stoner–wohlfarth model of ferromagnetism},}\ }\href {\doibase 10.1088/0143-0807/29/3/008} {\bibfield  {journal} {\bibinfo  {journal} {European Journal of Physics}\ }\textbf {\bibinfo {volume} {29}},\ \bibinfo {pages} {475} (\bibinfo {year} {2008})}\BibitemShut {NoStop}%
\bibitem [{\citenamefont {Cullity}\ and\ \citenamefont {Graham}(2008)}]{Cullity2008}%
  \BibitemOpen
  \bibfield  {author} {\bibinfo {author} {\bibfnamefont {B.~D.}\ \bibnamefont {Cullity}}\ and\ \bibinfo {author} {\bibfnamefont {C.~D.}\ \bibnamefont {Graham}},\ }\href@noop {} {\emph {\bibinfo {title} {Introduction to Magnetic Materials (2nd ed.)}}}\ (\bibinfo  {publisher} {Wiley-IEEE Press.},\ \bibinfo {year} {2008})\BibitemShut {NoStop}%
\bibitem [{\citenamefont {Adhikari}\ \emph {et~al.}(2022)\citenamefont {Adhikari}, \citenamefont {Liu}, \citenamefont {Wang}, \citenamefont {Hua}, \citenamefont {Liu}, \citenamefont {Lochner}, \citenamefont {Schlottmann}, \citenamefont {Yan}, \citenamefont {Zhao},\ and\ \citenamefont {Xiong}}]{adhikari2022interplay}%
  \BibitemOpen
  \bibfield  {author} {\bibinfo {author} {\bibfnamefont {Y.}~\bibnamefont {Adhikari}}, \bibinfo {author} {\bibfnamefont {T.}~\bibnamefont {Liu}}, \bibinfo {author} {\bibfnamefont {H.}~\bibnamefont {Wang}}, \bibinfo {author} {\bibfnamefont {Z.}~\bibnamefont {Hua}}, \bibinfo {author} {\bibfnamefont {H.}~\bibnamefont {Liu}}, \bibinfo {author} {\bibfnamefont {E.}~\bibnamefont {Lochner}}, \bibinfo {author} {\bibfnamefont {P.}~\bibnamefont {Schlottmann}}, \bibinfo {author} {\bibfnamefont {B.}~\bibnamefont {Yan}}, \bibinfo {author} {\bibfnamefont {J.}~\bibnamefont {Zhao}}, \ and\ \bibinfo {author} {\bibfnamefont {P.}~\bibnamefont {Xiong}},\ }\href@noop {} {\enquote {\bibinfo {title} {Interplay of structural chirality, electron spin and topological orbital in chiral molecular spin valves},}\ } (\bibinfo {year} {2022}),\ \Eprint {http://arxiv.org/abs/2209.08117} {arXiv:2209.08117 [cond-mat.mtrl-sci]} \BibitemShut {NoStop}%
\end{thebibliography}%

\end{document}